\documentclass[fleqn,usenatbib]{mnras}

\usepackage{newtxtext,newtxmath}

\usepackage[T1]{fontenc}
\usepackage{ae,aecompl}

\newcommand{\msun}{\mbox{$M_{\odot}$}}

\newcommand{\kms}{\ensuremath{\,\mathrm{km\ s}^{-1}}}

\usepackage{graphicx}	
\usepackage{amsmath}	
\hypersetup{linkcolor=red,citecolor=blue,filecolor=cyan,urlcolor=magenta}
\usepackage{natbib}
\usepackage[dvipsnames]{xcolor}

\title[Non-parametric Jeans Mass]
{Non-Parametric Spherical Jeans Mass Estimation with B-splines}

\author[Rehemtulla et al.]{
Nabeel Rehemtulla,$^{1,2,3}$\thanks{E-mail: nabeelr@umich.edu}
Monica Valluri,$^{1}$\thanks{E-mail: mvalluri@umich.edu}
and Eugene Vasiliev$^{4,5}$
\\
$^{1}$Department of Astronomy, University of Michigan, 1085 S.\ University Ave., Ann Arbor, MI 48109, USA\\
$^{2}$Department of Physics and Astronomy, Northwestern University, 2145 Sheridan Road, Evanston, IL 60208, USA\\
$^{3}$Center for Interdisciplinary Exploration and Research in Astrophysics (CIERA), 1800 Sherman, Evanston, IL 60201, USA\\
$^{4}${Institute of Astronomy, University of Cambridge, Madingley road, Cambridge, UK, CB3 0HA}\\
$^{5}${Lebedev Physical Institute, Leninsky prospekt 53, Moscow, Russia, 119991}\\
}

\date{Accepted XXX. Received YYY; in original form ZZZ}

\pubyear{2022}

\begin{document}
\label{firstpage}
\pagerange{\pageref{firstpage}--\pageref{lastpage}}
\maketitle

\begin{abstract}
Spherical Jeans modeling is widely used to estimate mass profiles of systems from star clusters to galactic stellar haloes to clusters of galaxies. It derives the cumulative mass profile, $M(<r)$, from kinematics of tracers of the potential under the assumptions of spherical symmetry and dynamical equilibrium. We consider the application of Jeans modeling to mapping the dark matter distribution in the outer reaches of the Milky Way using field halo stars. We present a novel non-parametric routine for solving the spherical Jeans equation by fitting B-splines to the velocity and density profiles of halo stars. While most implementations assume parametric forms for these profiles, B-splines provide non-parametric fitting curves with analytical derivatives. Our routine recovers the mass profiles of equilibrium systems with flattened haloes or a stellar disc and bulge excellently ($\lesssim 10\%$ error at most radii). Tests with non-equilibrium, Milky Way-like galaxies from the Latte suite of FIRE-2 simulations perform quite well ($\lesssim 15\%$ error for $r\lesssim 100$~kpc). We also create observationally motivated datasets for the Latte suite by imposing selection functions and errors  on phase space coordinates characteristic of Gaia and the DESI Milky Way Survey. The resulting imprecise and incomplete data require us to introduce an MCMC-based subroutine to obtain deconvolved density and velocity dispersion profiles from the tracer population. With these observational effects taken into account, the accuracy of the Jeans mass estimate remains at the level 20\% or better.
\end{abstract}
\begin{keywords}
-- galaxies: halo 
-- galaxies: kinematics and dynamics
-- galaxies: structure
\end{keywords}

\section{Introduction}
\label{sec:intro}
The European Space Agency's {\it Gaia} satellite \citep{perryman_etal_01, lindegren_etal_16, gaia_brown_16, gaia_brown_18}, launched in 2013, has released 3D positions and proper-motions for more than a billion Milky Way stars. In conjunction with line-of-sight velocities and chemical abundances for stars obtained with ground-based spectroscopic surveys (e.g. RAVE,\citealt{steinmetz_etal_06}; LAMOST, \citealt{LAMOST}; APOGEE, \citealt{APOGEE}; GALAH, \citealt{GALAH}; Gaia-ESO, \citealt{gaia-eso}, DESI \citealt{DESI-MWS_RNAAS_2020}),various components of the Milky Way's stellar halo are being characterized: field halo stars, stars in individual tidal streams, globular clusters, and satellites. The dynamics of these halo objects are important probes of the Milky Way's dark matter distribution and, in principle, allow us to determine several properties of the Milky Way's dark matter halo, most fundamentally its {\it mass density profile}. The mass of the Milky Way is a fundamental quantity of interest for comparisons with cosmological simulations,  but it is surprisingly poorly constrained. The availability of distances, radial velocities and even proper motions for huge numbers of individual stars, globular clusters and satellite galaxies has led to numerous efforts to determine the Milky Way halo parameters. Despite the availability of increasingly high quality data, the measurements of some basic properties of the Milky Way's dark matter halo have not converged. In Table 8 of \citet{bland_hawthorn_gerhard_16}, they summarize estimates for $M_{200}$ from 1999--2014 that range from $0.55-2.62\times 10^{12} M_\odot$. A more recent compilation by \citet{wang_20} considering only results obtained with Gaia DR2 data finds a similar range of values.  
Other methods using tidal streams, the velocities of satellite galaxies and field halo stars also give a large range of halo masses. This range of a factor of two in mass could be due to differences in the spatial distributions of tracers, systematic differences between methods, and differences in the treatment of errors and perturbations due to the LMC (which has only recently been accounted for in this context by \citealt{deason_etal_21} and \citealt{correa_21}).

The absence of high precision 6-D phase space coordinates (especially proper motions) for significant samples of halo kinematic tracers has led over the past two decades to the development of sophisticated techniques that incorporate the observational uncertainties via forward modeling. Starting with assumed parametric forms for the gravitational potential and distribution functions of tracers (star or globular clusters), these techniques impose observational errors and selection functions on the models to derive, often using Bayesian inference, the best fit estimates and their confidence intervals for a variety of potential parameters, including the mass of the halo. The most recent example of such distribution function modeling applied to Gaia (E)DR3 data \citep{deason_etal_21} is based on a sample of 665 halo stars (98\% with full 6D phase space data) in the distance range 50-100~kpc (including K giant stars, Blue Horizontal Branch (BHB) stars, RR Lyrae stars, and Blue Stragglers) and attempts to correct for the dynamical effect of the LMC.

We are on the verge of an era where the numbers of stars with full 6D phase space data is going to see a dramatic increase from hundreds to hundreds of thousands or even millions. The Dark Energy Spectroscopic Instrument (DESI) \citep{DESI_2016a, DESI_2016b, DESI-MWS_RNAAS_2020} is a 5000 fiber spectrograph on the Mayall 4-meter telescope at Kitt Peak National Observatory, Arizona, and is expected to obtain spectra of 5-8 million Gaia stars (down to $r=19.5$) 
in the next five years. In addition, the 4MOST \citep{4MOST} and WEAVE \citep{weave} surveys are similar multi-fiber spectrographic surveys which will together obtain spectra for millions of stars. These spectroscopic surveys will deliver line-of-sight velocities and spectro-photometric distances to stars out to 100~kpc, enabling -- for the first time -- the assembly of samples of stars with 6D phase space coordinates that are {\it orders of magnitude} larger than the best samples currently available.

In this work we implement a new version of one of the oldest and conceptually simplest dynamical modeling tools -- the spherical form of the Jeans equations \citep{Jeans_1915,Binney_1980,BT08} -- under the assumption that very soon we will have samples of $10^4-10^5$ halo stars out to $\sim 100$~kpc with  full 6D phase space coordinates. Our goal in this paper is to assess the effects of various factors on the derived cumulative mass profiles obtained with the spherical Jeans equation: (a) the underlying assumptions of the modeling approach: spherical symmetry and dynamical equilibrium; (b) realistic observational errors on proper motions, line-of-sight velocities and distances to halo stars and (c) the effects of limited survey volumes inherent to all surveys. 

Most previous implementations of the spherical Jeans equation, applied to the Milky Way halo either bin the data \citep{kafle_etal_18, wang_etal_18} and/or use analytic functions (e.g. power laws) to describe the density and velocity profiles and velocity anisotropy of tracer stars \citep{gnedin_etal_10}. Binning the data necessitates the computation of numerical derivatives which can be noisy \citep{kafle_etal_18}. For this reason, the use of analytic functions has been favored because they are relatively easy to implement and, in some cases, align well with theoretical predictions. For example, \citet{gnedin_etal_10} assign a power law relation to the tracer density profile of BHB stars beyond 25 kpc, which is broadly consistent with predictions for the  ``broken power-law'' density profile of the stellar halo from cosmological simulations \citep{Bullock_Johnston_2005, Johnston_etal_2008, Cooper_etal_2011}. Other studies use parametric fits to the velocity distribution and velocity anisotropy profile. These approaches have their limitations since parametric fitting curves may not adequately represent the true velocity or density profiles. Many studies construct parametric fits from binned data, which is sensitive to how bins are chosen, although discrete-kinematic Jeans models that avoid binning have also been developed \citep{Watkins_etal_2013}.  Since the velocity anisotropy profile, in particular, requires accurate determination of proper motions (which have only become available in large numbers in the past 3 years, thanks to Gaia), it has historically been common to assume either a constant value or a few functional forms \citep[e.g.][]{battaglia_05}, or to use functional forms motivated by cosmological simulations \citep{xue_etal_08,gnedin_etal_10}.  Only in recent years has it become possible to use the Gaia proper motions, sometimes even without line-of-sight velocities, to estimate the velocity dispersion tensor within $\sim 10$~kpc from the Sun and to constrain the Galactic potential from the axisymmetric Jeans equations \citep{Wegg_etal_2019, Nitschai_etal_2020}.

In the context of external dwarf spheroidal galaxies, sophisticated Bayesian methods have been developed to derive parametric fits to the tracer density, line-of-sight velocity distributions and velocity anisotropy profiles \citep[e.g. \textsc{GravSphere}, by][]{read_steger_2017}. \citet{Diakogiannis_bsplines_2017} have developed a hybrid approach and use B-splines to construct the radial velocity dispersion profile $\sigma_r^2$, while using a parametric mass model. They apply their method to line-of-sight velocities in the Fornax dwarf spheroidal galaxy. There have been several recent studies to assess how well the spherical Jeans equation performs on dwarf spheroidal galaxies when the various underlying assumptions are broken (e.g. spherical symmetry, dynamical equilibrium). For example, \citet{Evslin2017} use tracer particles drawn from N-body simulations to evaluate the fourth-order spherical Jeans equation when applied to dwarf spheroidal galaxies. \citet{genina_etal_2020} assess how \textsc{GravSphere} fares on mock dwarf spheroidal galaxies drawn from cosmological $\Lambda$CDM and SIDM simulations. \citet{El-Badry_jeans_etal_17} show that, using mock data sets from cosmological hydrodynamical simulations, dwarf galaxies with recent episodes of star formation and feedback are not in dynamical equilibrium and Jeans modeling results in an overestimate of the mass of the dwarf galaxy.  We do not discuss these methods for modeling external dwarf galaxies any further since the nature of the data, both in sample size and phase space dimensionality (typically only sky-positions and line-of-sight velocities for $<5000$ stars) are completely different from the situation considered in this work.

We present a novel B-spline based routine for performing spherical Jeans modeling for the Milky Way. Prior to applying it to real data, it is important to assess the modeling biases introduced by any spherical Jeans modeling code. This is pertinent because real potentials of Milky Way-like disc galaxies are not spherically symmetric nor are they in dynamical equilibrium. In addition, it is helpful to have a quantitative understanding the effect that breaking these assumptions has on the resulting mass profile. For this purpose we construct a diverse suite of mock datasets to validate our routine and quantify the errors in the derived cumulative mass estimate when the standard assumptions are broken. Although there have been several recent works focused on understanding the errors introduced by breaking various assumptions of the spherical Jeans equation by testing the method on simulations \citep[e.g.][]{El-Badry_jeans_etal_17,kafle_etal_18,wang_etal_18}, such tests are required to validate each code.  Consequently, the mock datasets in our suite include smooth, self-consistent equilibrium distribution functions generated using \textsc{Agama} \citep{vasiliev_19_agama} -- an efficient all-purpose galactic modeling package. We also generate mock stellar haloes from three galaxies of the Latte suite of FIRE-2 cosmological hydrodynamic zoom-in simulations. \citep{wetzel_etal_16,Sanderson_etal_ananke}. 

For the Latte galaxies, we also consider the effect of imposing survey selection functions and mock observational errors. For example, we exclude star particles outside the fiducial footprint of the Dark Energy Spectroscopic Instrument (DESI) \citep{DESI_2016a, DESI_2016b, DESI-MWS_RNAAS_2020}. Simultaneously, we impose mock observational errors on proper-motions, line of sight velocities and distances to tracer stars. Since selection functions and  measurement errors lead to poor Jeans mass estimates if not accounted for, we attempt to alleviate this by developing a Bayesian MCMC subroutine to deconvolve observational effects from the underlying data. We demonstrate the efficacy of this deconvolution process with our Latte-based mocks. Once DESI data are available, we will apply our Jeans routine to Gaia proper motions and distances and DESI line-of-sight velocities to construct a mass profile of the Milky Way out to $\sim 80-100$~kpc. 

The layout of this paper is as follows. Section~\ref{methods} gives a brief overview of spherical Jeans modeling and our B-spline implementation in the case of perfect and complete data (Sec.~\ref{sec:precise-complete}) and imprecise, incomplete data (Sec.~\ref{sec:obs-deconv}). Section~\ref{sec:mockdata} describes the various mock datasets we test the implementation on. Section~\ref{sec:results} describes the results of these tests.
We summarize and conclude in Section~\ref{sec:conclude}, and Appendix~\ref{sec:bsplines} gives an introduction to B-splines.

\section{Analysis Methods}
\label{methods}

The Jeans equations are  obtained from the collisionless Boltzmann equation by computing second order moments of the distribution function  \citep{BT08}. Under the assumption of spherical symmetry, they reduce to a single equation which can be used to derive the enclosed mass at any radius $M(<r)$, from which the full gravitational potential $\Phi$ can be obtained using $\Phi(r)=\frac{GM(<r)}{r}$. The form of the spherical Jeans equation we use here is given by
\begin{equation} 
    \label{sph_jeans_eq}
    M(<r) = 
    - \frac{r \overline{v_r^2}}{G}
    \bigg(
          \frac{\textrm{d ln }\rho}{\textrm{d ln }r} +
          \frac{\textrm{d ln }\overline{v_r^2}}{\textrm{d ln  }r} + 
          2\beta 
    \bigg).
\end{equation}
The velocity anisotropy parameter $\beta$ is defined as 
\begin{equation}
    \label{beta_eq}
    \beta = 1 - \frac{\overline{v_\theta^2} + \overline{v_\phi^2}}{2\overline{v_r^2}}.
\end{equation}
The terms $\overline{v_r^2}$, $\overline{v_{\phi}^2}$ and $\overline{v_{\theta}^2}$ are the means of the squares of the radial, azimuthal, and polar velocities respectively; $\rho$ is the number density of tracers (stars, globular clusters, etc.); $r$ is the spherical Galactocentric radius; $G$ is the gravitational constant. $\beta<0$ indicates a tangentially biased velocity distribution, $\beta=0$ indicates velocity isotropy, and $\beta>0$ indicates a radially biased velocity distribution. Note that in the above we use the full second moments $\overline{v_r^2}$, $\overline{v_{\phi}^2}$ and $\overline{v_{\theta}^2}$ rather than velocity dispersions $\sigma_r^2$, $\sigma_\phi^2$ and $\sigma_\theta^2$, since the former are more suitable for B-spline evaluation and these quantities are more relevant from the dynamical standpoint. However, in the cases with imprecise and incomplete data (Sec.~\ref{sec:obs-deconv}), we assume zero mean tangential streaming velocity ($\overline{v_\phi} = \overline{v_\theta}=0$) to reduce the number of free parameters, in which case $\overline{v_{\dots}^2} = \sigma_{\dots}^2$.

The spherical Jeans equation can be applied in two complementary regimes: (1) solving a ``dynamical inverse problem'' where one measures the dispersion of sky-plane and line-of-sight velocities directly from observations (with errors) and then ``inverts'' it to derive the radial velocity dispersion and anisotropy profiles (or assume $\beta$), and then infer  the potential or cumulative mass distribution (Eq.~\ref{sph_jeans_eq}); (2) ``forward modeling,'' on the other hand, assumes some parametric form for the gravitational potential and tracer distribution function and then computes the expected velocity dispersion and anisotropy profiles and convolves them with expected observational uncertainties and selection functions and compares with observations.  

Although dynamical inverse modeling was the original formulation of the method, inverse problems in general have difficulty dealing properly with errors and noise from small samples. Furthermore, it has been shown that in the context of the Milky Way halo, inferring the anisotropy profile from the line-of-sight velocity distribution results in a biased profile at Galactocentric distances beyond ${\sim}20$ kpc \citep{hattori_etal_17}. Most current applications therefore follow the second route because it is easier to account for observational errors. In this work, we assess how well the ``inverse modeling'' route behaves when full 6D phase space coordinate information is available (i.e. $\beta$ does not need to be inferred from $v_{\textrm{los}}$ alone), for samples that are orders of magnitude (${>}10^4$) larger than have previously been available. This method gives the potential (or rather the cumulative mass) more directly, but also suffers from errors more directly.

We split the analysis workflow into two stages. First, we reconstruct the radial profiles of tracer density $\rho(r)$ and components of second moment of velocity $\overline{v^2}_{r,\theta,\phi}$ (or the velocity dispersion $\sigma_{r,\theta,\phi}^2$, depending on the method) from the observed distribution of tracers. Then we substitute these functions into the Jeans equation (Eq.~\ref{sph_jeans_eq}) to infer the mass profile. The first stage could be used in a more general context, independently from the second stage, and is implemented in two variants, described in the following sections: neglecting the observational errors and incomplete spatial coverage, or accounting for them. In both variants, these functions are represented as B-splines -- a category of piecewise polynomials defined by an array of grid points (knots) and amplitudes of basis functions.

\subsection{The case of precise and complete measurements}
\label{sec:precise-complete}

The first approach is fairly simplistic and suitable only for situations in which the observational errors can be neglected and the spatial coverage is uniform. In this case, the logarithms of the density $\rho$ and the second moments of velocity $\overline{v_{i}^2}$ as functions of $\ln r$ are determined using out-of-the-box penalized spline fitting and log-density estimation routines implemented in \textsc{Agama}. The mathematical formalism is outlined in Appendix~\ref{sec:bsplines}. Specifically, we perform penalized spline regression to fit B-splines to $v_r^2$, $v_{\phi}^2$, $v_{\theta}^2$ and perform a penalized spline density estimate to calculate $\ln(\rho)$. An equally valid alternative is to construct an estimate of the kinetic energy density in each dimension $\ln(\rho\,v_{\dots}^2/2)$ and obtain the dispersions by dividing the kinetic energy density by the mass density.

There is no universally optimal procedure for determining the parameters of the knots: their number, spacing, minimum and maximum bounds. \citet{Diakogiannis_bsplines_2017} use a sophisticated but more computationally intensive evolutionary modeling algorithm to optimize and adapt their choice of knots when applying the \texttt{JEAnS} code to dwarf spheroidal galaxies. In contrast, we empirically determine our knot configurations, placing significant emphasis on keeping their parameters (minimum and maximum bound, and count) consistent over the variety of datasets to guard against overfitting.

In these fits, we use logarithmically spaced knots in radius (equally spaced in $\ln{r}$) because we find that this best captures the radial distribution of tracer particles. We adopt 6 knots for datasets described in Section~\ref{sec:agama_mocks} and 5 knots for the datasets described in Section~\ref{sec:latte_mocks} and \ref{sec:obs_mocks}. The minimum and maximum radial bound of the knots is similarly dataset dependent, but we have arranged them such that the knots are identical across all datasets presented in a given figure.

B-splines are an attractive choice for multiple reasons: chiefly, they provide analytical derivatives and non-parametric fits. The radial derivative terms in Eq.~\ref{sph_jeans_eq} on $\ln(\rho)$ and $\ln(\overline{v_r^2})$ are computed trivially by utilizing the B-splines' analytical derivatives, which are continuous across grid knots for B-splines of degree 2 and higher (we use degree 3 B-splines, which are equivalent to the familiar cubic splines). Further, the B-spline's non-parametric nature means that they do not restrict the velocity or density profiles to a specific shape as power-laws -- which are often used for Jeans modeling -- would.

\subsection{The case of imprecise and incomplete data: deconvolving observational effects from underlying distributions}
\label{sec:obs-deconv}

The second approach is designed to deal with the more complicated case when the observational errors are significant and the spatial coverage is non-uniform. We do this by forward modelling of the intrinsic density and velocity dispersion profiles, accounting for observational effects before comparing them to the actual data.

The distribution function of tracers is written in a factorized form,
\begin{equation}  \label{eq:df_factorized}
    f(\boldsymbol x, \boldsymbol v) = \rho(r)\;
    \prod_{i=1}^3 \mathcal N\big(v_i\;|\;{\overline{v_i}},\,\sigma^2_i(r)\big),
\end{equation}
where $\rho(r)$ is the density and $\mathcal N\big(v_i\;|\;\overline{v_i},\,\sigma^2_i\big)$ is the Gaussian velocity distribution for each of the three orthogonal velocity components in spherical coordinates (one radial and two tangential, the latter sharing the same velocity dispersion profile). As in the previous section, the logarithm of the density and the two components of the velocity dispersion $\sigma_{r,t}$ are represented by B-splines in $\ln r$ with fixed knots and free amplitudes varied during the fit. For simplicity, we assume zero mean streaming velocity ($\overline{v_r}=\overline{v_t}=0$), but in principle one could model it as another spline function of $\ln r$ with a few more free parameters. To account for the limited spatial coverage of the observed dataset, we normalize $\rho(r)$ to have unit integral over the survey footprint and over the range of heliocentric distances corresponding to the allowed range of apparent magnitudes (including the uncertainty in the distance modulus, as described in Section~4.3.1 of \citealt{hattori_etal_21}). This normalization needs to be computed with a relative error $\ll 1/N_\mathrm{stars}$ in order not to affect the likelihood, and we optimize the procedure for the specific case of the DESI survey footprint, which covers two spatially disjoint regions on the sky. The 3D integration across the sky coordinates and distance modulus is performed separately in each region with a suitable variable substitution transforming it into a unit cube.

The likelihood of the model against the observed dataset is given by convolving the distribution function (Eq. \ref{eq:df_factorized}) with observational uncertainties of each datapoint. \citet{hattori_etal_21} performed this convolution with the Monte Carlo method, splitting each star from the catalogue into $N_\mathrm{samples}\sim100$ points drawn from the joint error distribution of all relevant coordinates (distance, proper motion and line-of-sight velocity) and converting them to Galactocentric position and velocity; these samples are kept fixed in the course of MCMC analysis to minimize the impact of the Poisson noise on the likelihood. In their case, the model parameters specify both the potential and the distribution function of stars. To evaluate the likelihood of each model, the position and velocity of all samples are converted to actions computed in the model potential, and then the action-space distribution function of the model is averaged over all samples belonging to each star. In our case, the general approach is similar, but with two important differences. First, the likelihood of a model is determined by the phase-space (rather than action-space) distribution function (Eq.~\ref{eq:df_factorized}), and the corresponding potential of each model is computed from the Jeans equation at a later step. Second, we benefit from the fact that the error convolution can be performed analytically for all three velocity dimensions thanks to the Gaussian form of the distribution function, and use the Monte Carlo integration only to account for the distance uncertainty.
Specifically, we sample $N_\text{samples}=20$ points from each star's distance error distribution and convert the 3D coordinates of these samples into the Galactocentric radii $r_i^k$, where $i=1\dots N_\text{stars}$ enumerates stars and $k=1\dots N_\text{samples}$ iterates over samples for each star. These radii are kept fixed throughout the entire MCMC run. 
We also precompute auxiliary matrices for converting the two Galactocentric velocity dispersions $\sigma_{r,t}(r_i^k)$ into the dispersions of three heliocentric kinematic quantities -- two proper motion components $\sigma_{\mu_l;\,i}^k$,\, $\sigma_{\mu_b;\,i}^k$ and line-of-sight velocity $\sigma_{\text{los};\,i}^k$. Since the mean velocity in the model is assumed to be zero, the mean values of these heliocentric quantities are determined entirely by the Solar velocity and, in the case of proper motion, the distance to each sample, hence also precomputed in advance. Then for each choice of model parameters, we construct the splines for $\sigma_{r,t}(r)$, evaluate them at each sample's radius, convert into the Heliocentric coordinates, 
and add in quadrature the measurement uncertainties of each star $\epsilon_{\text{los};\,i}$,\, $\epsilon_{\mu_l;\,i}$,\, $\epsilon_{\mu_b;\,i}$ to obtain error-broadened dispersions of all three kinematic measurements (different for each sample). The likelihoods of the density, line-of-sight velocity, and both proper motion components are multiplied, averaged over all samples for each star, and the logarithms of these values are summed up to obtain the likelihood of the entire dataset:
\begin{equation}  \label{eq:likelihood_convolved}
\begin{array}{l}
\displaystyle \ln \mathcal{L} = \sum_{i=1}^{N_\text{points}} \ln \Bigg[ \frac{1}{N_\text{samples}}
    \sum_{k=1}^{N_\text{samples}} \rho(r_{i}^{k}) \;\times\\
\displaystyle \phantom{\ln \mathcal{L} =} \mathcal N\left( v_{\text{los};\,i} \;\big|\; \overline{v_{\text{los};\,i}},\,\left[\sigma_{\text{los};\,i}^k\right]^2 + \epsilon_{\text{los};\,i}^2 \right) \;\times \\
\displaystyle \phantom{\ln \mathcal{L} =} \mathcal N\left( \mu_{l;\,i} \;\;\,\Big|\; \overline{\mu_{l;\,i}^k},\;\;\,\left[\sigma_{\mu_l;\,i}^k\right]^2\: +\, \epsilon_{\mu_l;\,i}^2 \right) \;\times \\
\displaystyle \phantom{\ln \mathcal{L} =} \mathcal N\left( \mu_{b;\,i} \,\;\Big|\; \overline{\mu_{b;\,i}^k},\;\:\left[\sigma_{\mu_b;\,i}^k\right]^2 + \epsilon_{\mu_b;\,i}^2 \right)\;
    \Bigg].
\end{array}
\end{equation}

We first identify the best-fit model by maximizing the likelihood function, using the Nelder--Mead deterministic optimization algorithm. After this, we explore the parameter space with the MCMC search, using the \texttt{emcee} package \citep{emcee}. This gives the joint distribution of model parameters, and for each model in the chain, we compute the cumulative mass profile from the Jeans equation (Eq.~\ref{sph_jeans_eq}), as before, with the only difference that the second velocity moments are replaced by velocity dispersions.

We stress that even if we allowed for a nonzero mean velocity $\overline v$ and used the full second velocity moment $\overline{v^2}=\sigma^2+\overline v^2$ in the equations, this approach would not be equivalent to the one described in the previous section in the limit of complete sky coverage and negligible errors. The difference is subtle but might be important in some cases (e.g., a strongly anisotropic system): whereas in the first approach, we compute the kinetic energy directly, without assuming anything about the velocity distribution function, in the second approach we explicitly model it as a Gaussian function. In any case, we need to assume a certain functional form for the velocity distribution in order to perform the error convolution, and a Gaussian is by far the simplest and still fairly realistic case.

\section{Mock Data}
\label{sec:mockdata}

We now describe the various mock datasets used to validate our B-spline-based routine.
 
In Section~\ref{sec:agama_mocks} we describe the equilibrium mock datasets that were generated using \textsc{Agama}. These come in two categories: ``Halo-alone'' models (hereafter HA-models) and ``Halo-Disc-Bulge'' models (hereafter HDB-models). The purpose of these tests is to verify the accuracy of the routine and to assess the effects of realistic deviations from sphericity and, in the case of the HDB-models, the effect of contamination of halo tracers by disc particles.
 
In Section~\ref{sec:latte_mocks} we describe more realistic mock datasets generated from star particles drawn from three galaxies (\textit{m12f, m12i}, and \textit{m12m}) of the Latte suite of FIRE-2 cosmological hydrodynamic zoom-in simulations \citep{wetzel_etal_16}. These models introduce significantly more complexity and realism, including the effects of halo substructure like streams and clustered halo stars, and disequilibrium arising from tidal interactions with nearby satellites. They also present the challenge of accurately separating disc and halo stars.

In Section~\ref{sec:obs_mocks} we describe how we impose observational selection functions and errors onto the mocks described in \ref{sec:latte_mocks}. The selection functions we consider are for the fiducial DESI footprint and survey magnitude limits of the Gaia and DESI surveys. The mock errors are imposed on proper motions, line-of-sight velocities, and distance moduli. These mocks also introduce a dependence on the choice of solar position, which we vary between three locations in the disc.


\subsection{Self-consistent equilibrium mock datasets}
\label{sec:agama_mocks}

The \textsc{Agama} dynamics package is used to generate self-consistent equilibrium distribution functions with all particles having the same mass. In practice, the halo tracers provided to the Jeans routine from these datasets are dark matter particles, but we assume in these simplified models that halo stars would be similarly distributed. 

The HA-models contain only a spheroidal halo with axis ratio $q$ varying from $q=1.0$ (spherical) to $q=0.6$. \textsc{Agama} uses the double-power-law distribution function in action space from \citet{posti_2015} in which we tailor the mixing coefficients to produce the desired oblate axisymmetric shape. The mixing coefficients in the linear combination of actions in the distribution function are responsible for both the flattening and the velocity anisotropy, and it may not be straightforward to construct a model that follows a particular spatial density profile precisely. We determine the parameters of the distribution function (power-law indices, mixing coefficients, etc.) that approximately reproduce the \citet{hernquist_90} density profile. We then tune the mixing coefficients to create a desired amount of flattening, which we quantify by the axis ratio of the inertia tensor, as described in \citet[their method E1]{zemp_2011}. The models are fully specified by the distribution function, and the corresponding potential-density pair is determined via an iterative procedure \citep{binney_14}. Once generated, the halo mass and scale radius are scaled to $M_{\textrm{halo}}=1.3 \times 10^{12} M_\odot$ and $R_{\textrm{halo}}=25 \textrm{ kpc}$, roughly matching the Galactic virial mass and halo density break radius quoted in \cite{bland_hawthorn_gerhard_16}.  

The HDB-models have a spherical halo, a disc, and a spherical bulge. The HDB's initial halo density function adds an exponential cutoff to the HA density function but scales to the same mass $M_{\textrm{halo}}=1.3 \times 10^{12} M_\odot$ with the same scale radius $R_{\textrm{halo}}=25 \textrm{ kpc}$. The double-exponential disc is a combination of the thin and thick disc parameters quoted in \cite{bland_hawthorn_gerhard_16}. Summing the thin and thick disc's stellar mass yields $M_{\textrm{disc}} = 4.1 \times 10^{10} M_\odot$ and a mass weighted average of the radial scalelengths and scaleheights for the thin and thick discs yields $R_{\textrm{disc}}=2.5 \textrm{ kpc}$ and $h_{\textrm{disc}}\approx375 \textrm{ pc}$. 
The spherical bulge follows the \texttt{Spheroid} density profile with parameters taken from \citet{mcmillan_17}: total mass $M_{\textrm{bulge}}=8.9 \times 10^{9} M_\odot$, scale radius $R_{\textrm{bulge}}=0.075 \textrm{ kpc}$, outer cutoff radius $R_\text{cut}=2.1 \textrm{ kpc}$, and power-law indices $(\alpha,\beta,\gamma)=(1,1.8,0)$. 

While we create variations of the HA-models with different axis ratios, the halo in the HDB-model remains spherical. However, we make additional HDB-models with other alterations. In one variant, we inject a random sample containing $1/4$ of the system's disc particles into the halo tracer population to test the routine's sensitivity to contamination of the halo tracer population with disc particles. In another mock, we assess the effect of a radially varying velocity distribution for halo particles by creating a halo particle distribution with the Cuddeford--Osipkov--Merritt anisotropy profile \citep{Osipkov_1979, Merritt_1985_AJ, Merritt_1985_MNRAS, cuddeford_91}, using $\beta(r=0)=0.2$ and anisotropy radius $r_a=100$ kpc.

The HA-models and HDB-models are generated with 300\,000 particles in their haloes. The HDB-model with disc contamination has an additional 40\,000 particles used as input, which come from the 160\,000 particles comprising its disc.

\subsection{Mock datasets from cosmological hydrodynamic simulations}
\label{sec:latte_mocks}
The Latte suite of FIRE-2 cosmological simulations are performed using mesh-free hydrodynamics from the GIZMO code \citep{Hopkins_2015} and the Feedback in Realistic Environments model 2 (FIRE-2) \citep{Hopkins_etal_2014,wetzel_etal_16}. The Latte galaxies are a collection of simulated Milky Way-mass galaxies at uniquely high resolution that are well suited for generating  mock Milky Way-like datasets. We use the redshift $z=0$ public snapshots of the \textit{m12f, m12i}, and \textit{m12m} galaxies released with the \textit{Ananke} synthetic Gaia catalog \citep{Sanderson_etal_ananke}. We create datasets for each galaxy by selecting only the star particles with metallicities  $\lbrack\mathrm{M}/\mathrm{H}\rbrack<-1.5$ and ages $t_{\textrm{age}}>8$~Gyr. Note that we do not impose a geometric cut to exclude disc stars since this complicates the estimation of the density profile, we assume that our metallicity and age selection largely suffices to exclude disc stars. This is a proxy for selecting metal-poor accreted halo stars since the public FIRE snapshots do not contain information about whether stars were accreted from satellites or formed \textit{in situ} in the host galaxy. Since information about whether a star was accreted is also not available \textit{a priori} in real data, this provides an additional level of realism to our mock datasets. We calculate the $\lbrack\mathrm{M}/\mathrm{H}\rbrack$ ratio and the hydrogen mass fraction ($X$) from the mass fractions of He ($Y$) and metals ($Z$) provided in the public Latte data as follows:

\begin{eqnarray}
X & = & 1-(Y+Z)\\
\lbrack\mathrm{M}/\mathrm{H}\rbrack & = & \log(Z/X)-\log(Z_\odot/X_\odot)
\end{eqnarray}

These mocks and the mocks in Section \ref{sec:obs_mocks} operate under the simplification considering each Latte star particle to be a single tracer star. We consider our tracers to be RR-Lyrae stars (RRLs) because they serve as standard candles. Each metal poor halo star particle (${\sim}7070~\msun$) can be considered to represent the entire stellar population of approximately a single star cluster. 
%
It is generally considered reasonable to assume that on average a star cluster of this mass would contain one RRL.\footnote{Milky Way Globular clusters, whose stellar populations are similar in age and metallicity range to the stellar halo, have masses $\sim 5\times 10^4-10^6$\msun and contain on average $\sim 1.7\times 10^{-4}$ RRlyrae per \msun. Alternatively assuming that the stellar halo has a mass of $\sim 10^9$\msun and given that the ``clean sample'' of  Gaia RR Lyrae stars from \citet{iorio_belokurov_19} contains  $N=93\,345$ RRlyrae, the are roughly $9.3\times 10^{-5} \sim 10^{-4}$ RRL/\msun.}
After selecting tracers by metallicity and age, there are approximately 200\,000 remaining particles in each dataset (\textit{m12f}: 187\,457, \textit{m12i}: 229\,347, \textit{m12m}: 217\,286).

We note in passing that we chose not to use mock stars from the {\it Ananke} mock Gaia catalogs \citep{Sanderson_etal_ananke} of the same Latte galaxies for two reasons. First,  although {\it Ananke} produces stellar populations and phase space distributions of stars with uncertainties, the currently public catalogs are only for Gaia DR2 uncertainties, while we wish to generate mock catalogs for Gaia (E)DR3 and DR5. Second, the process employed to create the {\it Ananke} catalogs ``explodes'' each star particle from the simulation into a stellar population. While the stellar population particles preserve the local phase space density around the original star particle, this population is isotropic in velocity space, i.e. it does not preserving the local velocity anisotropy in the vicinity of the original star particle. Since the cumulative mass estimate (Eq.~\ref{sph_jeans_eq} \& Eq.~\ref{beta_eq}) depends explicitly on the velocity anisotropy of the stellar tracers, the mock stars from Ananke, with their isotropic velocity distribution would bias our results.

\begin{figure*}
    \begin{center}
        \includegraphics[trim=0 0 0 0,clip,height=8.6cm,keepaspectratio]{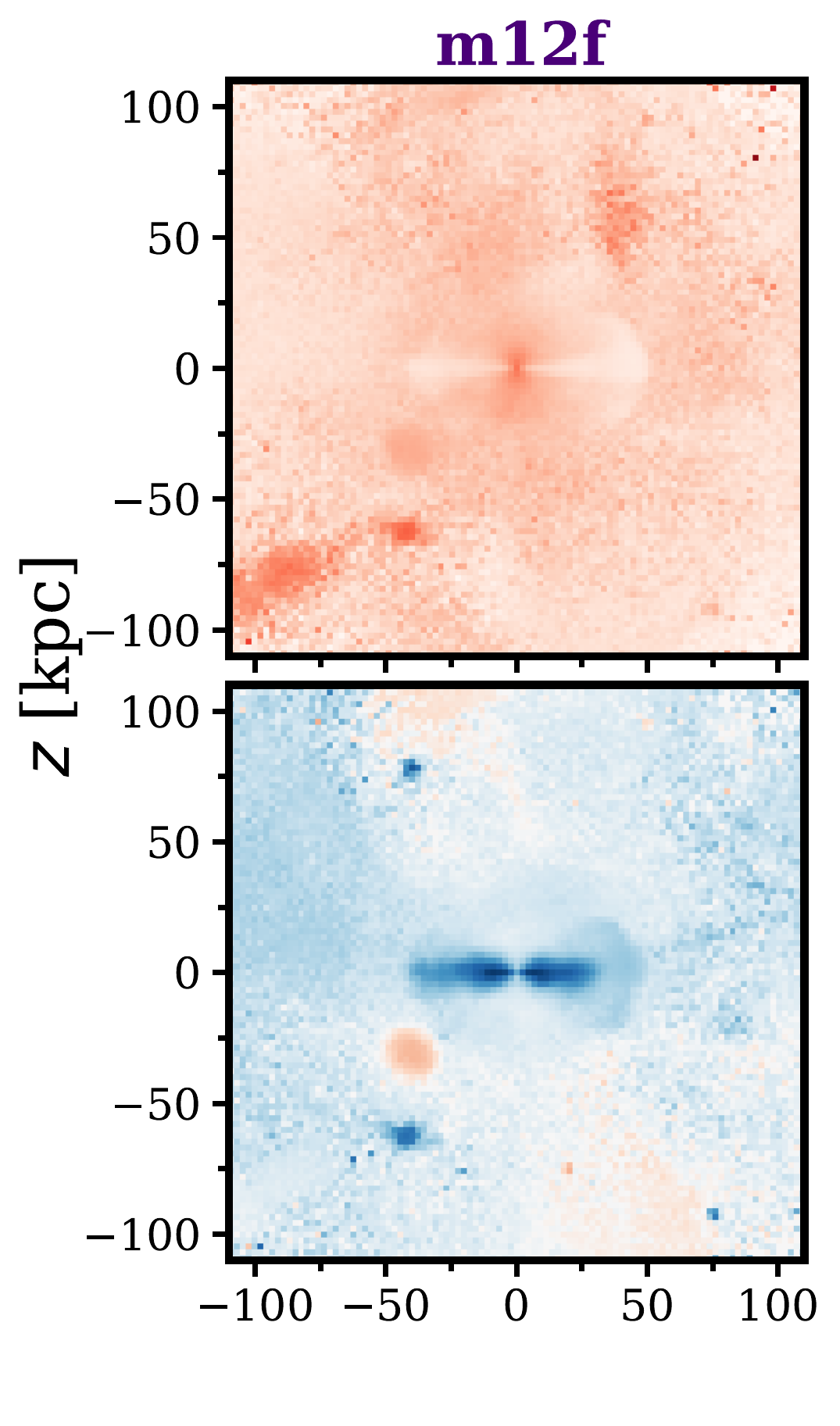}
        \includegraphics[trim=0 0 0 0,clip,height=8.6cm,keepaspectratio]{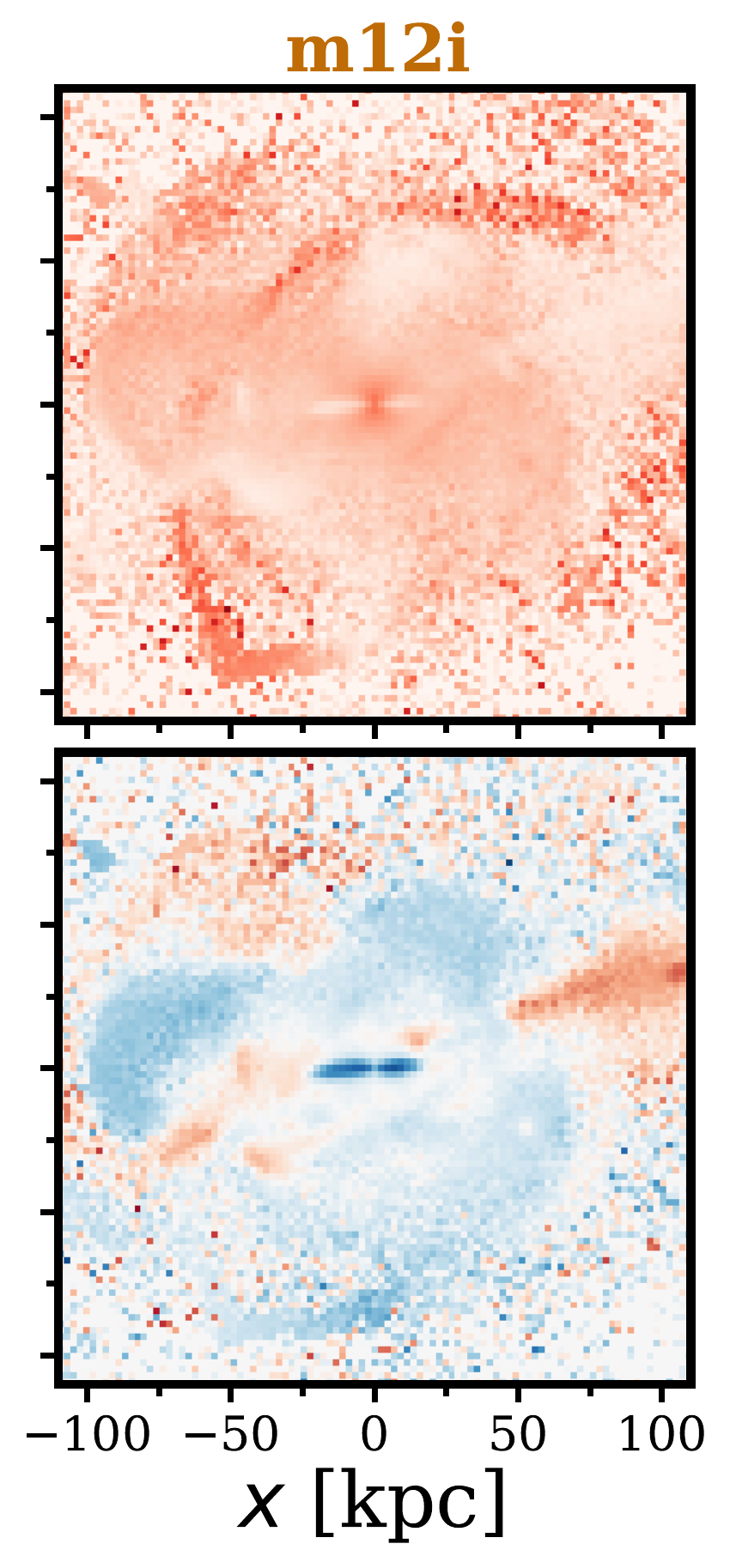}
        \includegraphics[trim=0 0 0 0,clip,height=8.6cm,keepaspectratio]{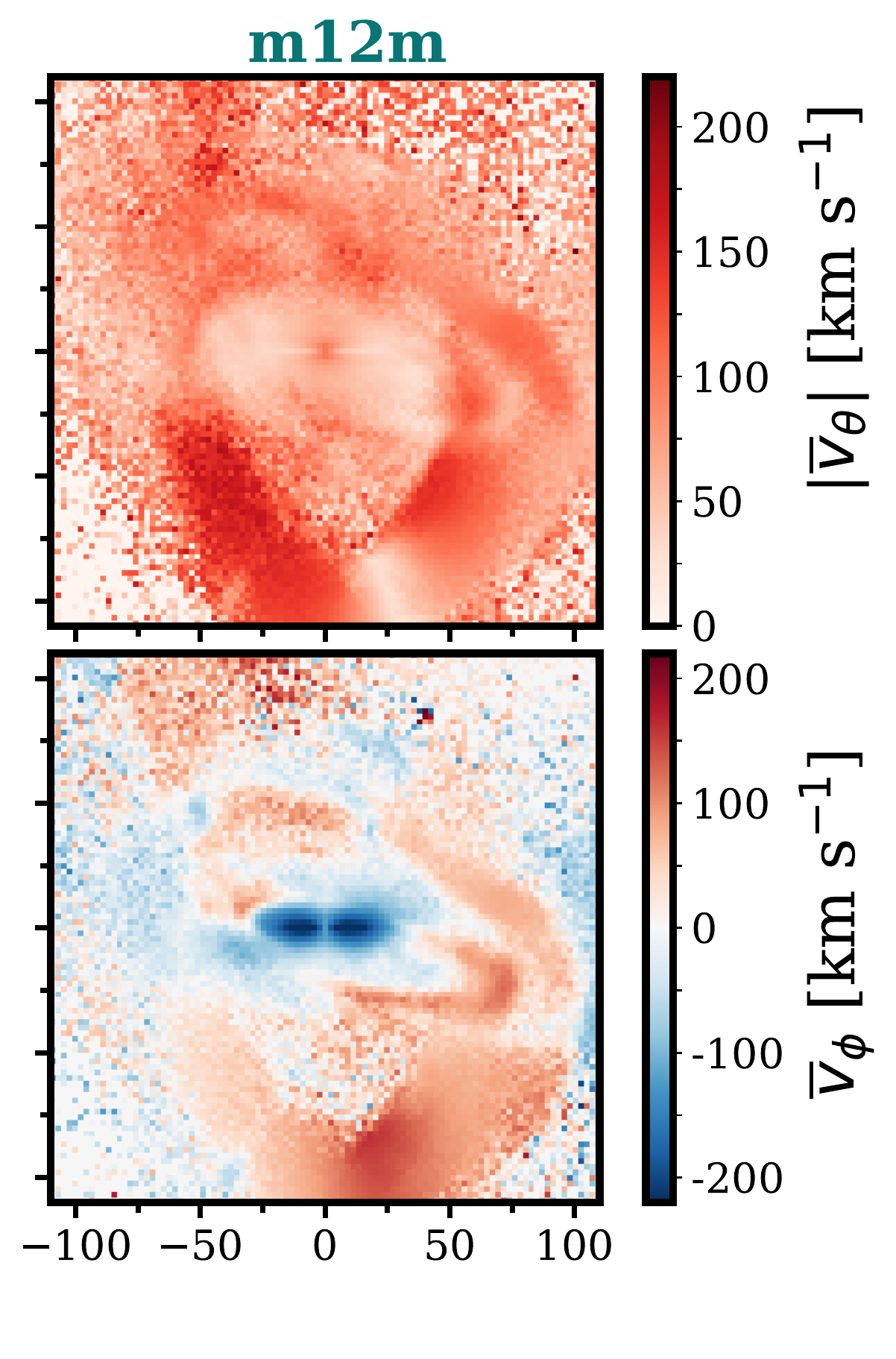}
        \caption{2-dimensional heatmaps colored by $|\overline{v}_\theta|$ and $\overline{v}_\phi$ for the Latte galaxies \textit{m12f} (left column), \textit{m12i} (center column), and \textit{m12m} (right column). All three show significant streaming motion and substructure, evidence of deviations from dynamical equilibrium, violating the assumptions underlying the spherical Jeans equation.}
        \label{fig:heatmaps}
    \end{center}
\end{figure*}

The three Latte galaxies, while all Milky Way mass, have varying amounts of halo substructure in the form of tidal streams and shells and a fairly broad range of satellite properties. Their haloes are also in different states of dynamical equilibrium, have a range of realistic morphologies, have experienced a varied assembly history \citep{Garrison-Kimmel_etal_2018}, and have realistic satellite populations \citep{wetzel_etal_16}. In addition, these galaxies have halo stars with moderate amounts of tangential streaming motions (azimuthal $\overline{v_\phi}$ and polar $\overline{v_\theta}$). Figure~\ref{fig:heatmaps} shows heatmaps of star particles in these three galaxies from the present-day (redshift zero) snapshot, color coded by $|\overline{v_\theta}|$ (upper row) and $\overline{v_\phi}$ (lower row). All three  galaxies, show significant streaming motions and substructure but \textit{m12m} has the most dramatic streaming motion, so we expect our Jeans estimation to perform the worst on the mock datasets derived from it.

In addition to substructure in the form of streams, shells and over-densities, there is recent observational evidence that the inner part of the Milky Way, including the Sun, is moving with respect to the outer stellar halo as a result of on-going tidal interaction with the Large Magellanic Cloud (LMC) \citep{petersen_penarrubia_2021,erkal_etal_2021} and that the stellar halo density distribution shows a distinct polarization pattern also likely caused by the LMC \citep{conroy_etal_2021}. \citet{erkal_etal_20} and \citet{deason_etal_21} demonstrated that these perturbations lead to an overestimate of the Milky Way mass profile with traditional dynamical modelling methods relying on the equilibrium assumption, and \citet{correa_21} presented a method for compensating these perturbations in mass modelling. Although the main effect of the LMC is kinematical, by coincidence, the magnitude of Milky Way mass overestimate roughly matches the mass of the LMC itself, if we were to add it to the virial mass of our Galaxy, as noted in \citet{li_etal_20}. In this paper we do not consider the effects of significant perturbations, such as that arising from tidal interaction with the LMC.

\subsection{Adding observationally motivated selection functions and errors}
\label{sec:obs_mocks}


One must choose a solar position or a Local Standard of Rest (LSR) before imposing observationally motivated selection functions (SFs) or errors on mock halo stars from simulated galaxies. As part of the construction of the \textit{Ananke} synthetic Gaia catalog \citep{Sanderson_etal_ananke}, the Latte collaboration provided three LSRs with precomputed solar velocities for each of the three galaxies \textit{m12f, m12i}, and \textit{m12m}. The first LSR position, LSR0, places the sun at $(x,y,z)=(0, 8.2, 0)\textrm{ kpc}$ and the other two, LSR1 and LSR2, are located at the same Galactocentric radius, in the galactic midplane, and rotated $120^\circ$ in either direction from LSR0. We transform the coordinates of the star particles to the equatorial and Galactic coordinate systems with \textsc{Agama} when applying the following selection functions and errors. 

We start by imposing a sky-selection function similar to the fiducial DESI footprint with a declination cut $-35^\circ \leq \delta \leq 90^\circ$ and a galactic latitude cut $30^\circ \leq |b| \leq 90^\circ$. We also impose a magnitude selection function in the Gaia $G$-band ($16.0<G<20.7$), the lower bound determined by the DESI Milky Way Survey (MWS) \citep{DESI-MWS_RNAAS_2020} and the upper bound by Gaia's expected performance\footnote{\url{https://www.cosmos.esa.int/web/gaia/science-performance}}. 
The tracer star particles from the Latte galaxies represent metal-poor RR Lyrae stars (RRL), for which the distance modulus $\mathcal D$ can be computed from the apparent magnitude $G$, assuming the absolute $G$-band magnitude $M_G=0.58$ with an intrinsic scatter 0.24 mag, corresponding to about a 10\% distance error \citep{iorio_belokurov_19}. We compute the distance modulus for our mock stars as follows:
\begin{eqnarray}
\label{eq:distmod}\mathcal{D}=5*\textrm{log}(d/\textrm{kpc})+10\\
\label{eq:G_app}G = M_G + \mathcal{D}
\end{eqnarray}

\begin{figure*}
    \centering
        \includegraphics[width=0.8\paperwidth]{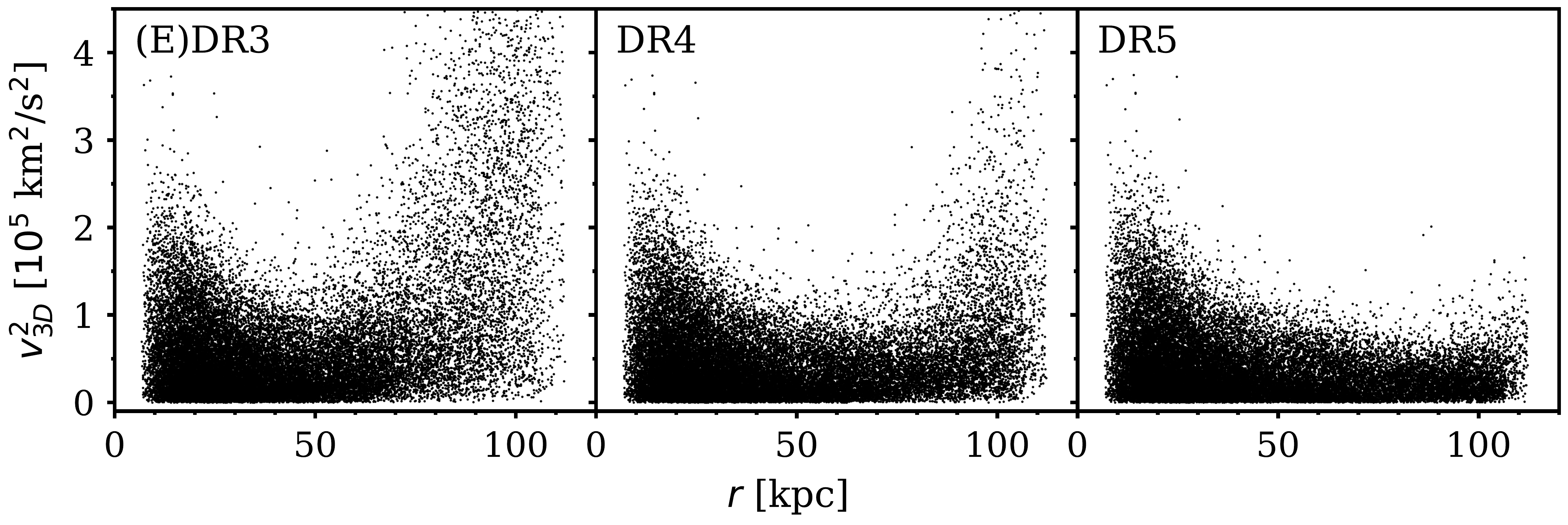}
        \caption{3D velocities squared ($v_{\textrm{3D}}^2$) of tracers in mock dataset for \textit{m12f} at LSR0 with errors on proper motion ($\mu_{\alpha*}, \mu_{\delta}$), distance modulus ($\mathcal{D}$), and line-of-sight velocity ($v_\textrm{los}$). From \textit{left} to \textit{right}: $v_{\textrm{3D}}^2$ values computed assuming proper-motion errors ($\sigma_{\mu_{\alpha*}}$, $\sigma_{\mu_{\delta}}$) expected from Gaia data releases (E)DR3, DR4, and DR5 respectively.  Since $\sigma_{\mu_{\alpha*}}$ and $\sigma_{\mu_{\delta}}$ increase with increasing $G$-band apparent magnitude, stars at large Galactocentric radii $r$ have more uncertain tangential velocities resulting in larger $v_{\textrm{3D}}^2$ values.  The number of apparently unbound stars dramatically decreases with progressive Gaia data releases as the expected $\sigma_{\mu_{\alpha*}}$ and $\sigma_{\mu_{\delta}}$ decrease.}
        \label{fig:gaia_drs}
\end{figure*}
The mock errors are randomly generated from independent Gaussian distributions on each quantity for each star particle with standard deviations determined by characteristic or expected errors for Gaia and DESI observations. We use a very conservative line-of-sight velocity error for the DESI Milky Way Survey $\sigma_{v_\textrm{los}}=10$~\kms \citep{DESI-MWS_RNAAS_2020}, which is reasonable for RRLyrae which are pulsating stars, although the $v_{\textrm{los}}$ errors are expected to be 2-5~\kms for other tracer populations such as luminous red giants. The Gaia proper motion errors ($\sigma_{\mu_{\alpha*}}$, $\sigma_{\mu_{\delta}}$) come from \texttt{PyGaia}\footnote{PyGaia: \url{https://github.com/agabrown/PyGaia} written by A.G. Brown}. \texttt{PyGaia} allows for calculations of errors from Gaia (E)DR3, and expected errors from DR4 (containing the first 5\,${}^1{\mskip -5mu/\mskip -3mu}_2$ years of observations) and DR5 (containing up to 10 years of observations). The uncertainty in distance modulus $\sigma_\mathcal{D}$ is the same for all particles and is caused by the intrinsic scatter in the RRL period--luminosity relation. We add distance modulus errors first, because the proper motion errors are magnitude dependent. Lastly, we assume the right ascension ($\alpha*$) and the declination ($\delta$) have no error.

Figure~\ref{fig:gaia_drs} shows the 3D velocities squared of tracers in \textit{m12f} after imposing the listed observational selection functions and errors at LSR0 over three Gaia data releases. Because the Gaia proper motion errors increase with increasing $G$, tracers at large Galactocentric radii tend to have their velocities dramatically boosted, often above the expected escape velocity curve. These extremely high velocity stars will cause enclosed mass estimate from Jeans modeling to blow up at large radii if not corrected for by our more advanced modeling described in Section~\ref{sec:obs-deconv}. This effect makes the datasets described in this section practically unusable without deconvolution especially when generated with proper motion errors corresponding to earlier Gaia releases: (E)DR3 \& DR4. The forward modeling method described in Sec.~\ref{sec:obs-deconv} also allows us to correct for the volume removed by the DESI footprint when calculating density. 

The mock datasets described here are LSR-dependent, so imposing these SFs and errors produces three datasets for each Latte galaxy -- one for each of the three LSRs -- and have roughly 20\,000--30\,000 remaining tracers each. Further, for each Gaia release -- (E)DR3, DR4, DR5 -- we make a separate mock dataset, totaling to 27 possible mocks. We only show results for (E)DR3 and DR5 as the DR4 results lie between the other two as expected.

\section{Results}
\label{sec:results}

Each of the mock datasets described in Section~\ref{sec:mockdata} is provided as input to our routine, the B-spline functions for $\ln\rho(r)$, $\overline{v^2}_{r,\phi,\theta}(r)$
are generated, and the resulting $M(<r)$ is computed using Eq.~\ref{sph_jeans_eq}. This estimated mass profile is compared with the system's true mass profile over the relevant radial range. The true mass profile is computed for the entire radial extent of the system by cumulatively summing the masses of all particles (stars, gas, and dark matter where relevant) within a given Galactocentric radius.

The estimated mass profile from each mock is judged for accuracy by computing the percent error relative to the corresponding true mass profile. Figures~\ref{fig:HA_fig}-\ref{fig:latte_fig} and \ref{fig:deconv_fig} show the estimated mass profiles (solid lines), their percentage errors (lower panels), and the true mass profiles (dashed lines) as a function of Galactocentric radius for a collection of mock datasets. The mocks described in each subsection of Section~\ref{sec:mockdata} correspond to a matching subsection in Section~\ref{sec:results}.

\subsection{Initial tests with self-consistent equilibrium mocks}
\label{sec:agama_results}
\begin{figure}
    \begin{center}
        \includegraphics[width=1.0\columnwidth]{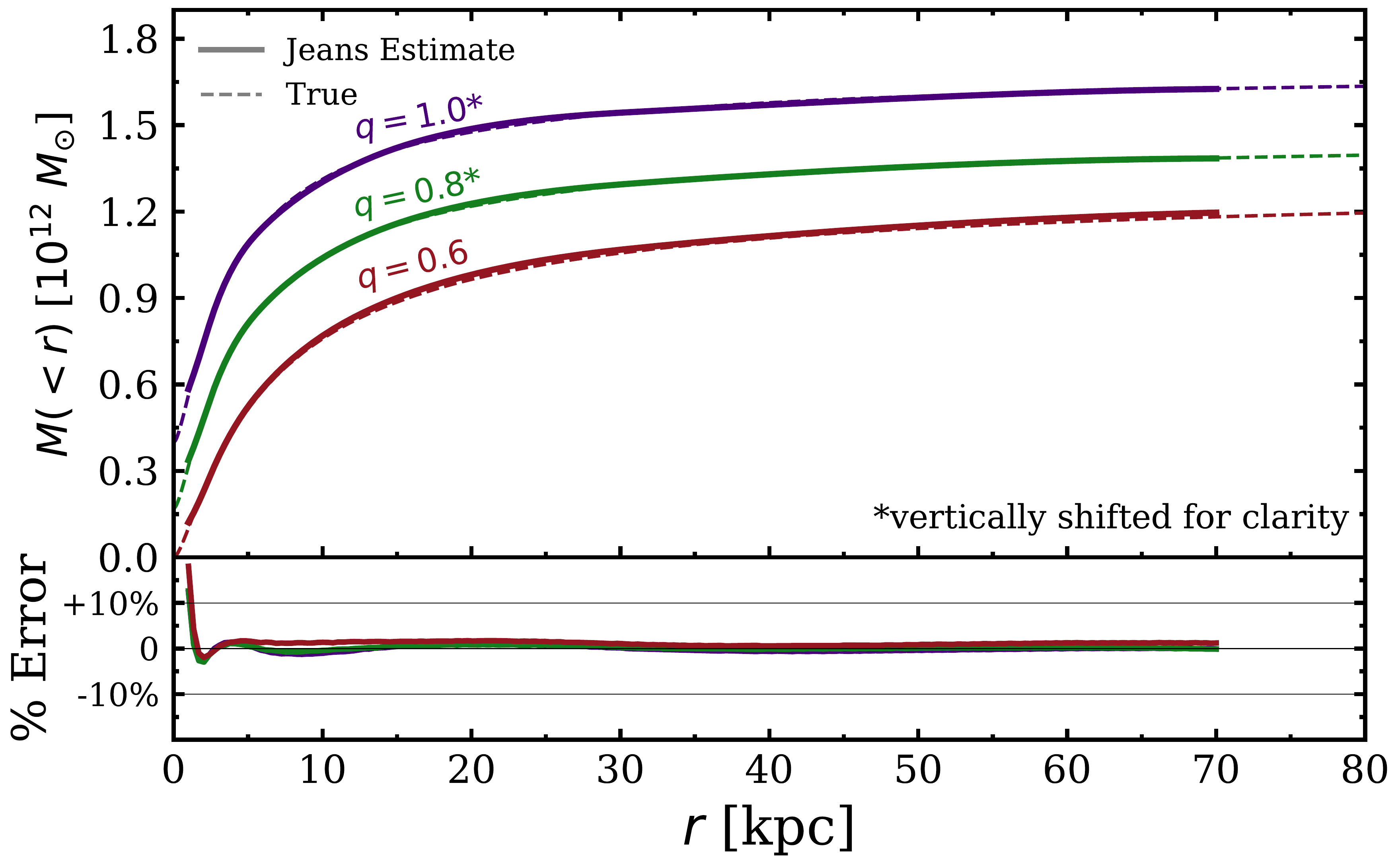}
        \caption{Dashed curves show true cumulative mass distributions and solid curves show the spherical Jeans estimate of the same. The lower panel shows the \% error in the mass estimate. Mass profiles of three ``Halo-alone'' (HA) models with flattening quantified by the axis ratio $q=1.0$ (spherical)$, 0.8,$ and $0.6$ (oblate axisymmetric). Increasing flattening causes minor fluctuations of no more than $4\%$ from the true mass at most radii. The larger error at small radii is most likely due to small numbers of tracers. These suggest that sphericity is not absolutely required for very good estimations with spherical Jeans modeling.}
        \label{fig:HA_fig}
    \end{center}
\end{figure}

In this section we describe the results from tests with the self-consistent equilibrium mock datasets described in Section~\ref{sec:agama_mocks}. Figure~\ref{fig:HA_fig} shows the results for the three HA-models, whose B-spline fits are each run with $N_{\textrm{knots}}=6$ knots logarithmically spaced between $r_{\rm min}=1$ kpc and $r_{\rm max}=70$  kpc. Here, we also remove particles beyond the bounds of the knots when performing the spline fits. The blip in error within $r<5$ kpc is merely an artifact of small number statistics at small radii. Aside from this, the progression of flattening from $q=1.0$ to $q=0.6$ only has a very small effect on the accuracy of the estimation. The mass profile from the $q=0.6$ mock oscillates around the true profile and is noticeably less stable than that of the $q=1.0$ and $q=0.8$ mocks. For all three mocks however, the error is less than 4\% beyond $r=5$~kpc. The very small error in the estimations on the $q=0.8$ and $q=0.6$ mocks shows that breaking the sphericity assumption does not necessarily mean the Jeans estimate will be poor. We also run tests on HA models with $q=0.9$ and $q=0.7$, which perform in-line with the results shown here.

\begin{figure}
    \begin{center}
        \includegraphics[width=1.0\columnwidth]{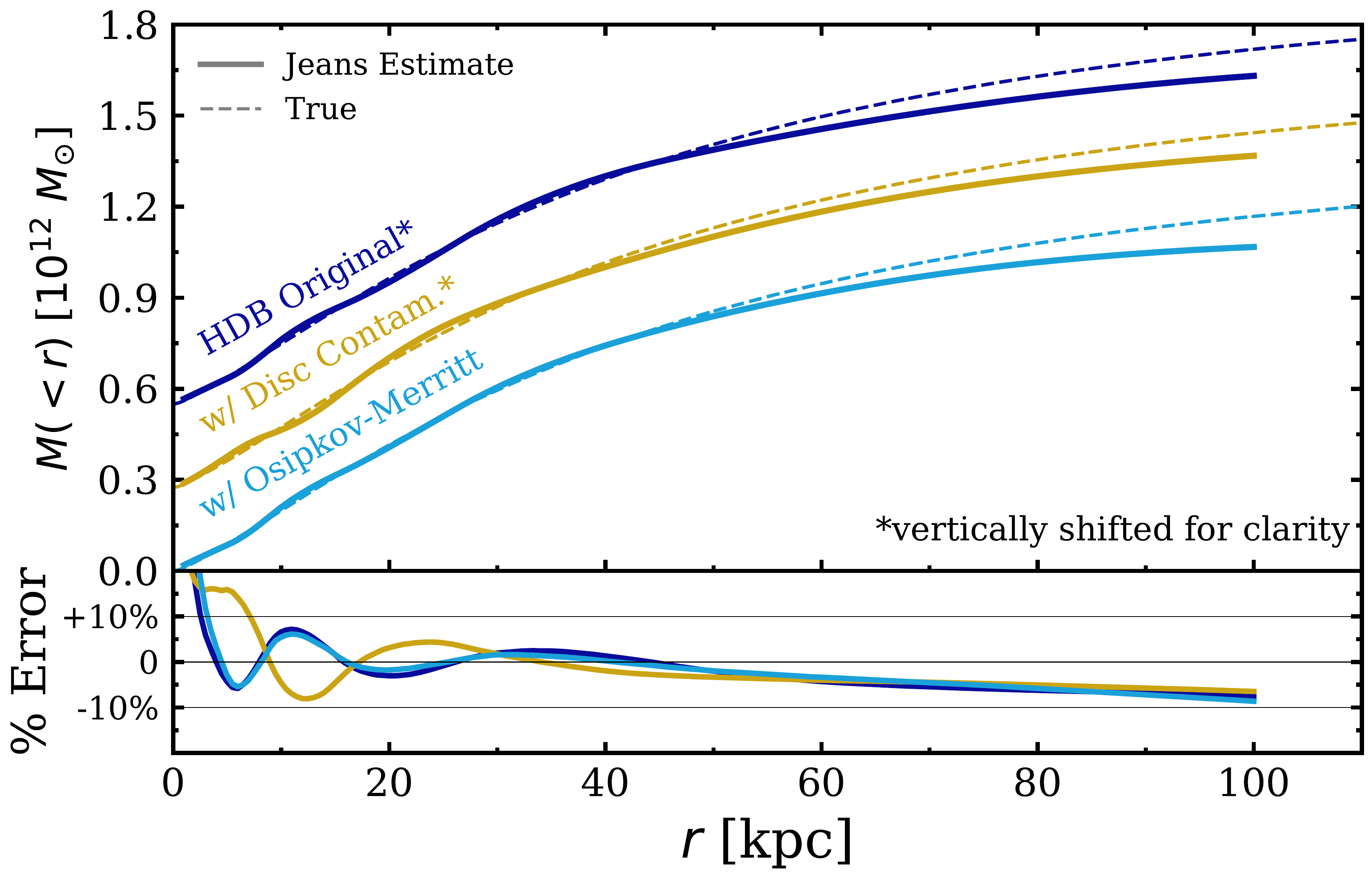}
        \caption{As in Fig.~\ref{fig:HA_fig}, dashed curves show true distributions and solid curves show our Jeans estimates for mass profiles of the ``Halo-Disc-Bulge'' (HDB) models including variants with contamination from disc particles and a radially varying Cuddeford--Osipkov--Merrit velocity anisotropy profile. All variants perform very well:  error ${\leq}10\%$ at most radii. The inclusion of a disc breaks the assumption of sphericity but has little effect on the accuracy, even when the tracer population is contaminated with disc particles (yellow curve). A radially varying  anisotropy profile also imparts very little error beyond $r=30$ kpc where the disc ends.}
        \label{fig:HDB_fig}
    \end{center}
\end{figure}

In MW-like galaxies, the sphericity assumption is also broken by the presence of a disc. Figure~\ref{fig:HDB_fig} shows the results of tests with the original HDB-model and its variants. For these models, the B-spline fits are all performed with $N_{\textrm{knots}}=6$ knots logarithmically spaced between $r_{\rm min}=5$ kpc and $r_{\rm max}=80$ kpc. The log-scaled spline functions in log-radius are linearly extrapolated beyond the grid endpoints, corresponding to a power-law extrapolated density and kinetic energy profiles. The error in the estimation on the HDB-models is within $10\%$ beyond $r=10$ kpc. Notably, the significant deviations between the estimation on the original HDB model and its variants are confined to roughly $r<30$ kpc. Beyond that, the three mass profiles are quite similar. The differences at small radii between the disc contamination variant and the others are expected as the injection of rapidly orbiting and spatially flattened disc particles are significant changes to that region of the system.

\subsection{Early tests with metal-poor particles from cosmological simulations}
\label{sec:latte_results}

\begin{figure}
    \begin{center}
        \includegraphics[width=1.0\columnwidth]{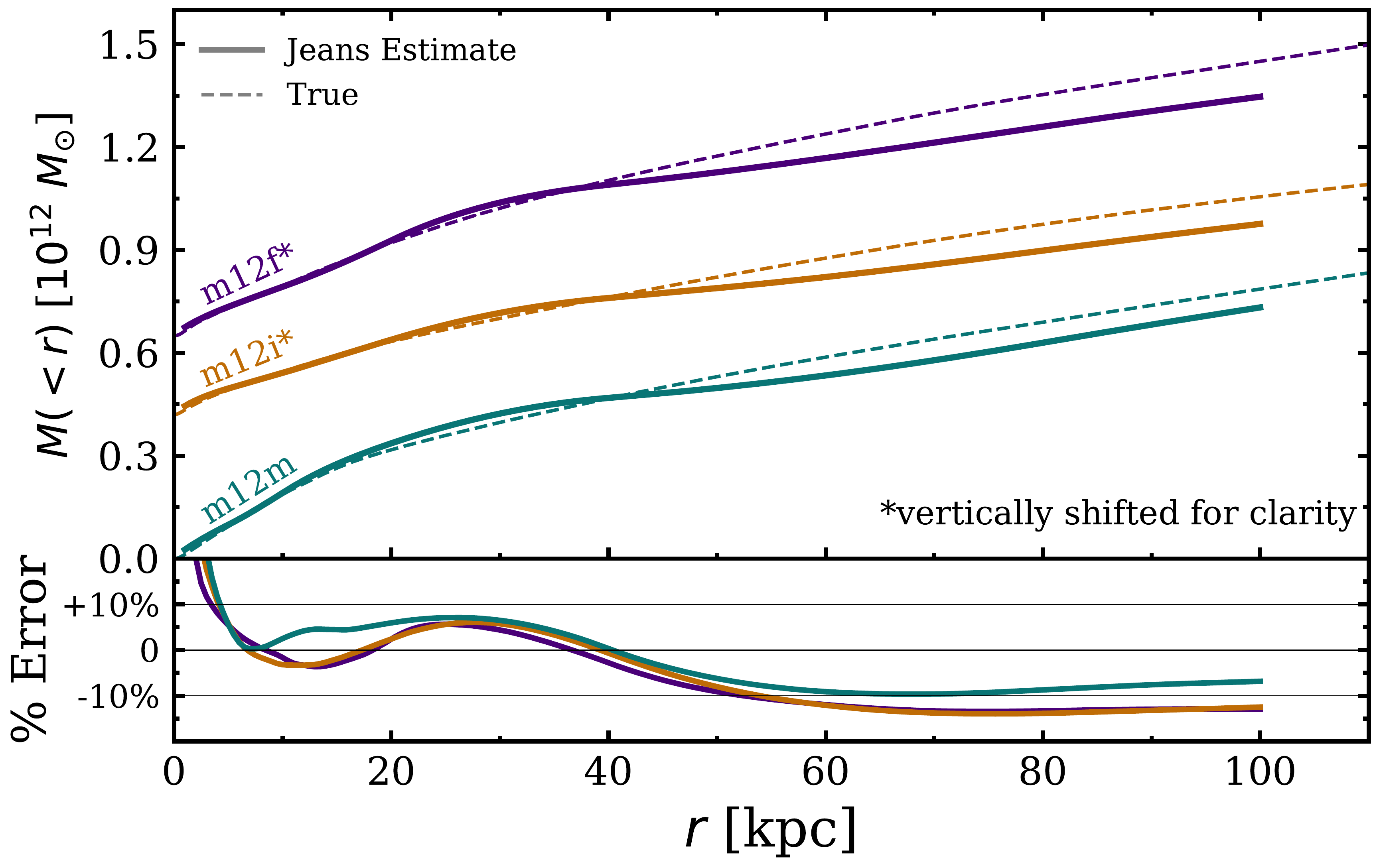}
        \caption{Mass profiles of mock datasets generated using three galaxies from the Latte FIRE-2 cosmological simulations: \textit{m12f} (purple), \textit{m12i} (orange), and \textit{m12m} (teal). For each, we emulate a stellar halo RR Lyrae population by selecting star particles with $[\textrm{M}/\textrm{H}]<-1.5$ and $t_{\textrm{age}}>8$~Gyr. All three resulting mass profiles are within $15\%$ of the truth beyond $r=5$~kpc. These tests with advanced cosmological simulations illustrate our B-spline routine's ability to smooth out disequilibrium like tidal streams, halo substructure, and tangential streaming motion to recover the underlying mass distribution quite well.}
        \label{fig:latte_fig}
    \end{center}
\end{figure}

In this section we describe tests of our Jeans modeling routine with mock halo stars drawn from galaxies belonging to the Latte suite of FIRE-2 cosmological simulations: {\it m12f, m12i}, and {\it  m12m}. The datasets we generate using these three galaxies select a population of metal-poor, old star particles with $[\textrm{M}/\textrm{H}]<-1.5$ and $t_{\textrm{age}}>8$~Gyr meant to represent a stellar halo RR Lyrae population. The B-spline fits on these datasets are each run with $N_{\textrm{knots}}=5$ knots logarithmically spaced between $r_{\rm min}=5$ kpc and $r_{\rm max}=80$ kpc. We reduce the number of knots from the tests with equilibrium mocks (Section~\ref{sec:agama_results}) to increase the smoothness in the B-spline fits. Figure~\ref{fig:latte_fig} shows the results of the tests for these three galaxies when all metal poor stars are selected. The accuracy with which our Jeans modeling routine recovers the input mass distribution is, of course, worse for these mocks than for the idealized equilibrium mocks. The resulting mass profiles all have very similar shapes with inflection points in the same radial regions. They are all within 15\% error nearly everywhere and often much less. Like the HDB-models, they uniformly exhibit underestimates at roughly $r>40$~kpc in either set. Here, we only show estimates out to $r=100$~kpc for ease of comparison with the observationally motivated results in the next section (the DESI and Gaia magnitude limits will yield too few stars and large velocity errors beyond this distance). Nonetheless, we maintain a very similar level of accuracy for our cumulative mass estimate ($\lesssim15\%$ error) when we use error-free Latte data with $(r_\textrm{min},r_\textrm{max},N_\textrm{knots})=(5 \textrm{ kpc},100 \textrm{ kpc},5)$ evaluated out to $r=200$~kpc. We also obtain a similar level of accuracy ($\lesssim 20\%$ error) when using variants of these mocks with different metallicity thresholds ([M/H] < -1.0, -2.0) but the same age threshold.

In the following section, it is important to keep in mind that the three Latte galaxies used here have triaxial dark matter haloes, show significant halo substructure (shells and tidal streams), show varying amounts of tangential streaming (see $\overline{v}_\phi$ of Fig.~\ref{fig:heatmaps}), and have satellites within 100~kpc. Despite these various sources of disequilibrium and spherical asymmetry, Figure~\ref{fig:latte_fig} shows that, with error-free and complete data, the cumulative mass profiles are still well recovered. Further, we observe that the introduction of these sources of disequilibrium are more detrimental to the accuracy than the introduction of spherical asymmetries on our equilibrium models shown in Sec.~\ref{sec:agama_results}. The distribution function fitting performed on mock stellar halo catalogues in \citet{wang_etal_15} made the same assumptions that we do here, spherical symmetry and dynamical equilibrium, and similarly found that errors from disequilibrium are dominant over those from asymmetry.

\subsection{Tests with observational errors and selection functions on cosmological simulations}
\label{sec:obs_results}

\begin{figure*}
    \begin{center}
        \includegraphics[trim=0 65 0 0,clip,width=8cm,keepaspectratio]{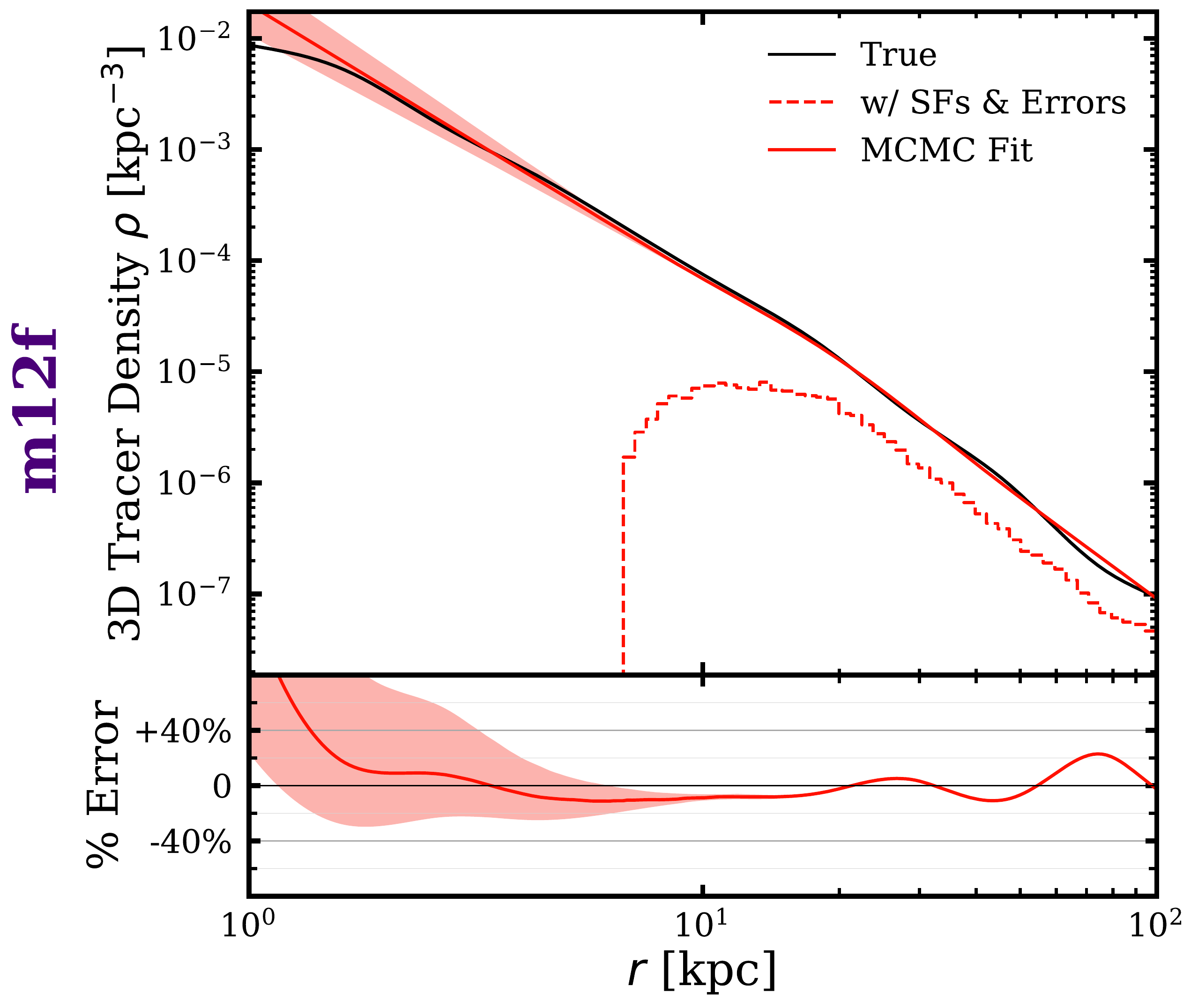}
        \includegraphics[trim=0 65 0 0,clip,width=7.5cm,keepaspectratio]{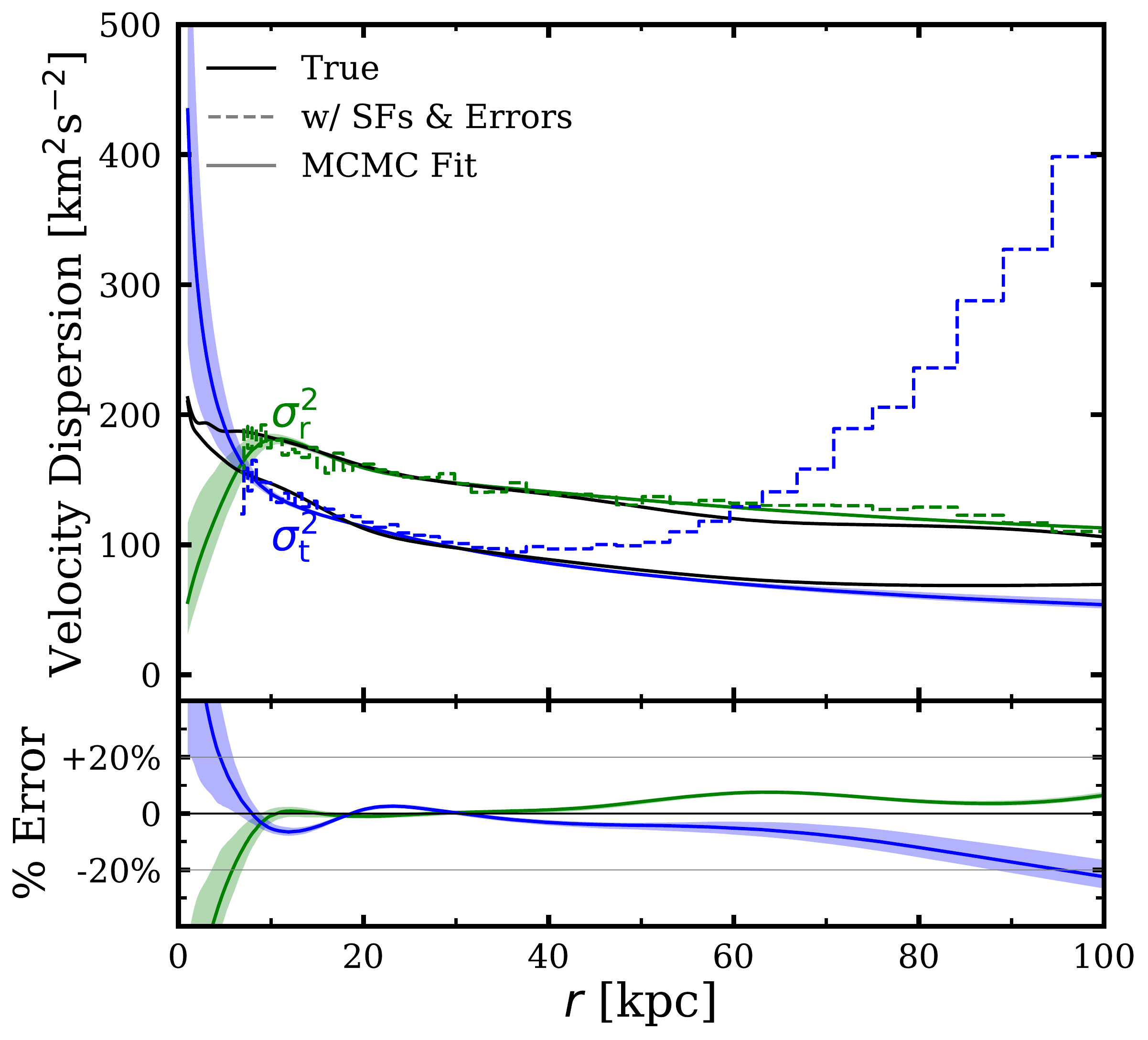}
        \includegraphics[trim=0 65 0 0,clip,width=8cm,keepaspectratio]{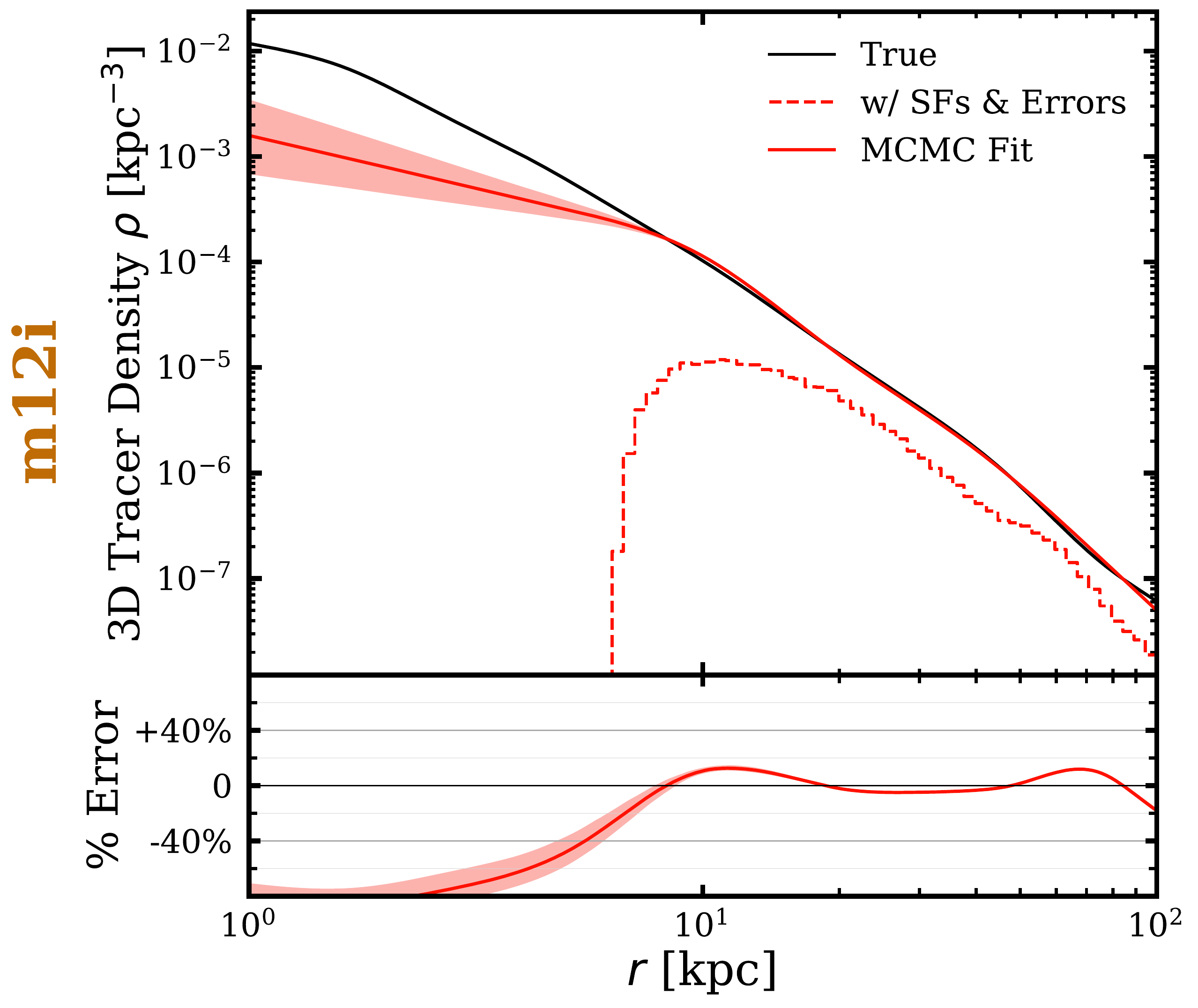}
        \includegraphics[trim=0 65 0 0,clip,width=7.5cm,keepaspectratio]{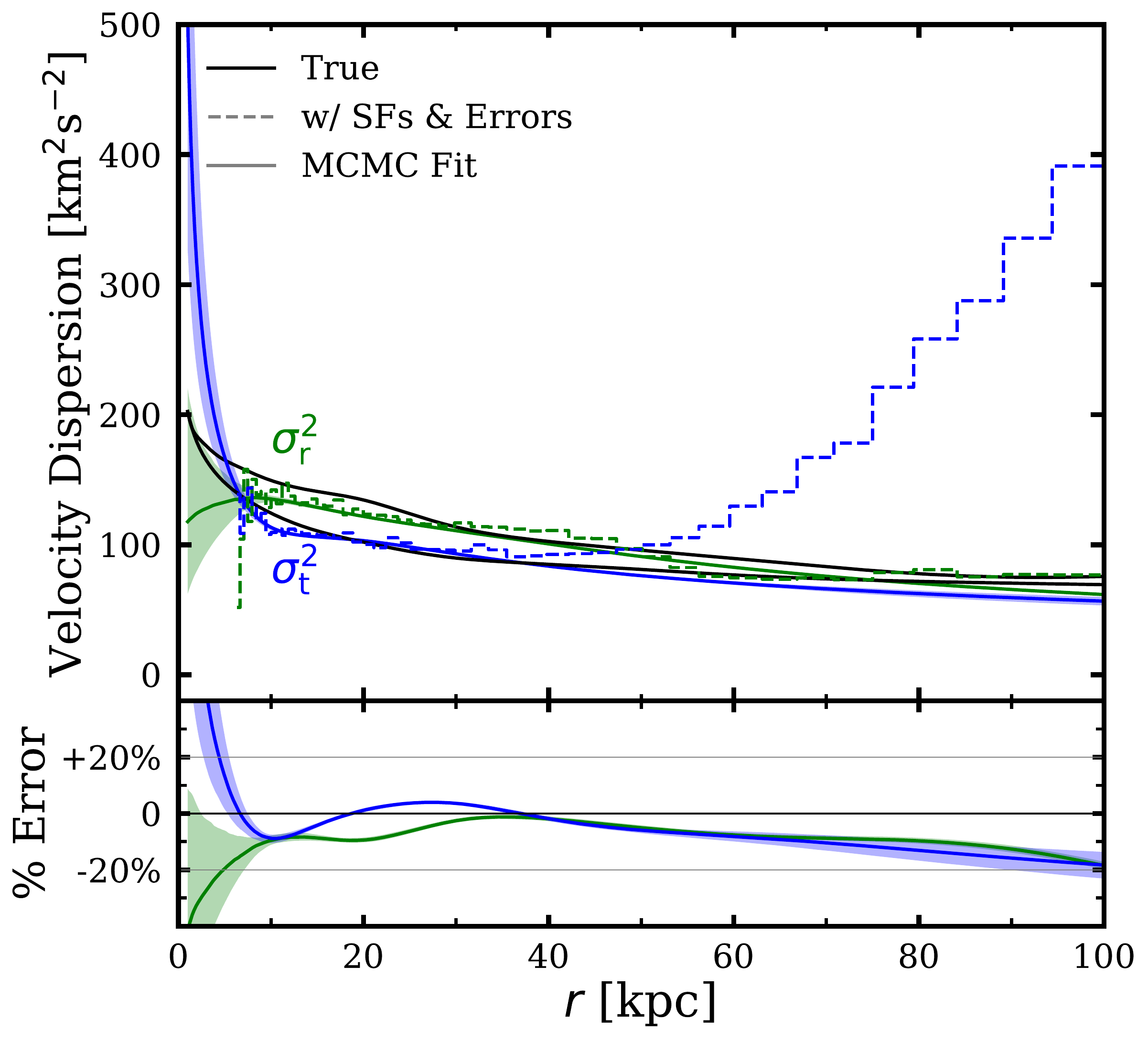}
        \includegraphics[trim=3 0 0 0,clip, width=7.96cm]{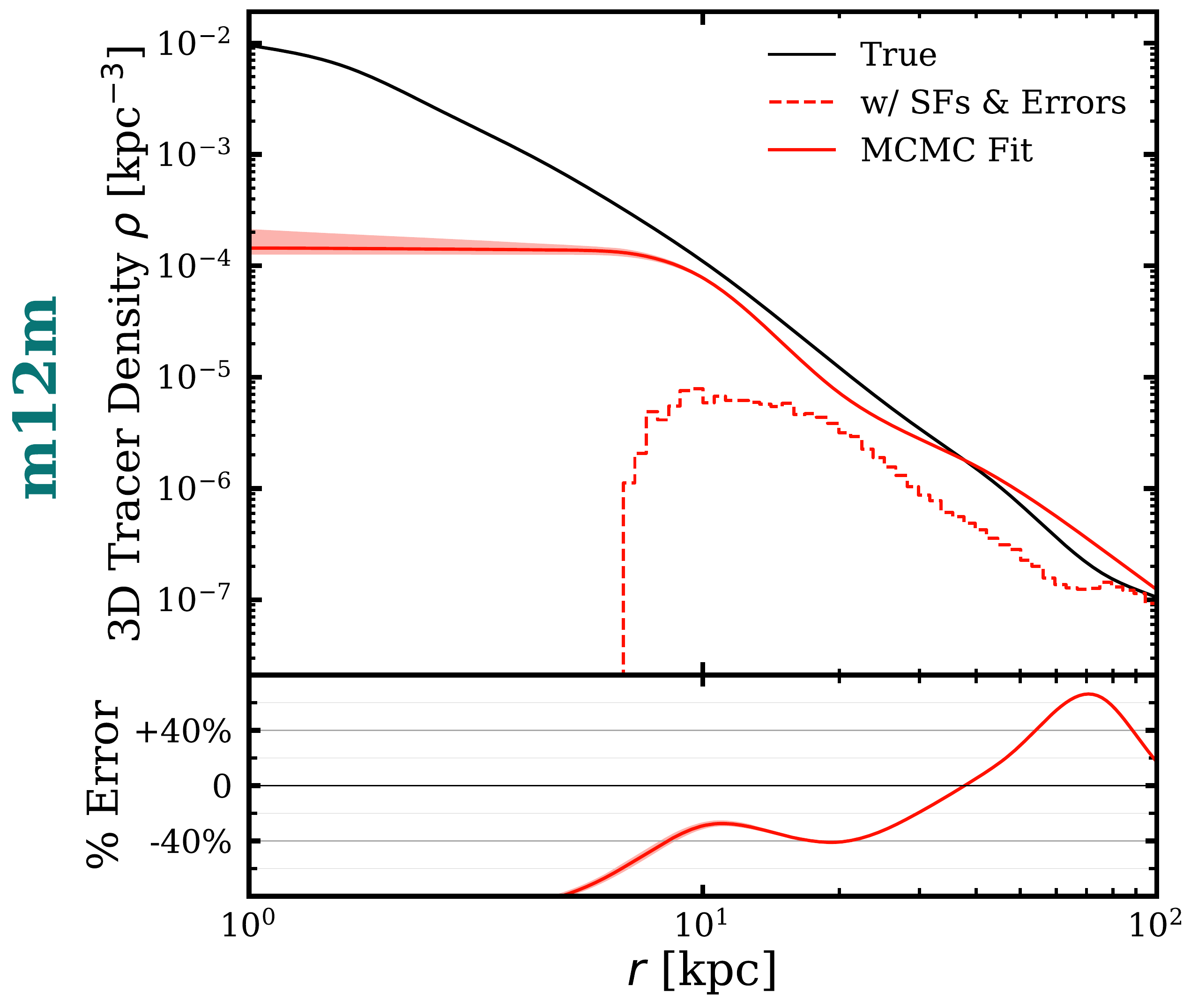}
        \includegraphics[trim=0 0 10 0,clip, width=7.4cm]{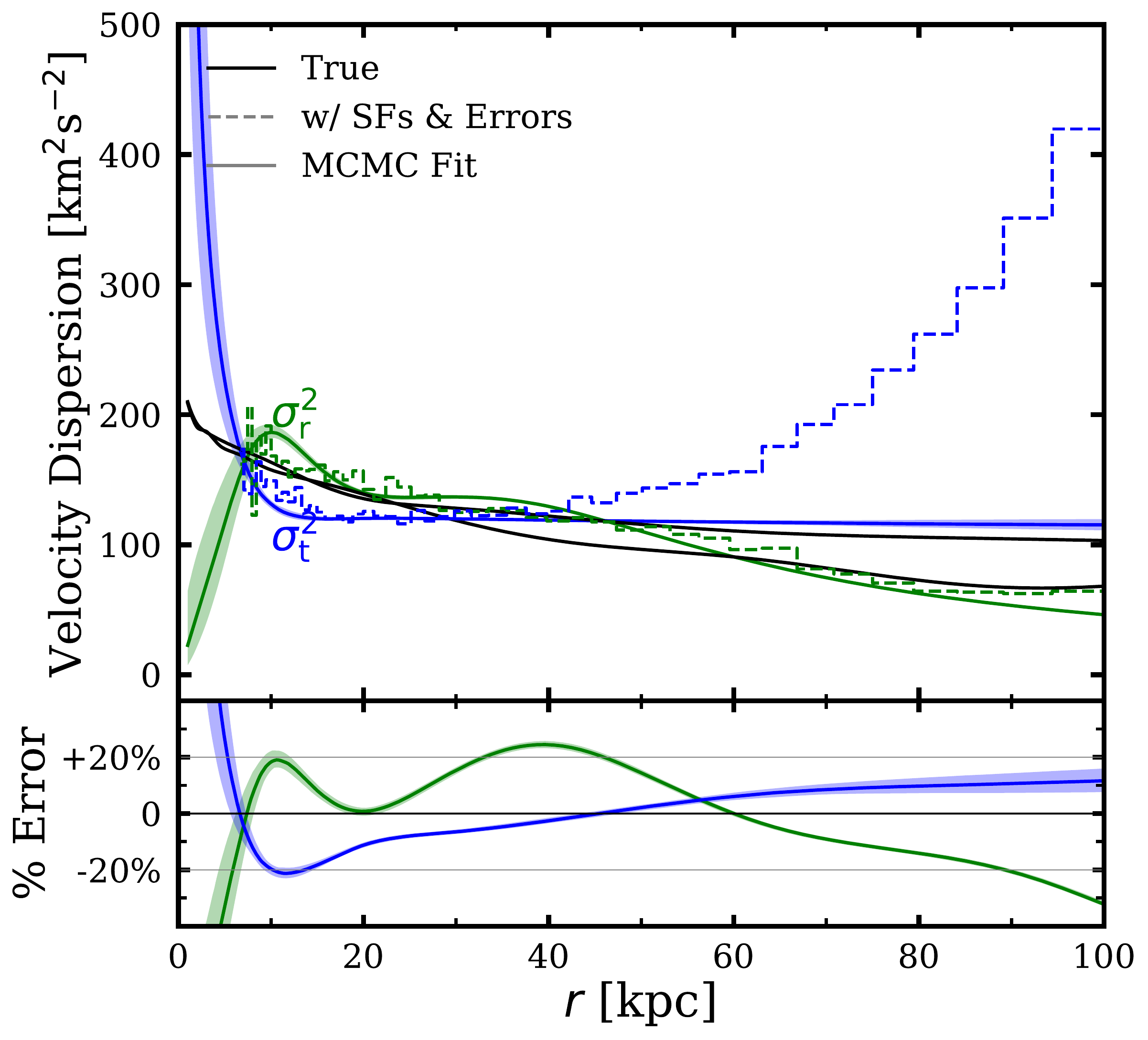}
        \caption{Tracer number density (red, $\rho$), radial velocity dispersion (green, $\sigma_r^2$), and tangential velocity dispersion (blue, $\sigma_t^2$) of Latte galaxies \textit{m12f} (top row), \textit{m12i} (middle row), and \textit{m12m} (bottom row). For each galaxy (each row), the solid black lines are the true profiles made with all old metal-poor star particles; the dashed colored lines are the profiles obtained after imposing observational selection functions and errors; the solid colored lines are the MCMC fits computed by the deconvolution routine and are used for the Jeans estimates in Section~\ref{sec:obs_results} and Figure~\ref{fig:deconv_fig}. The smaller panels show the percent error of the deconvolved distributions from the true profiles. Errors on proper motions are characteristic of Gaia (E)DR3; errors on line-of-sight velocity $v_\textrm{los}$ are 10~\kms; errors on distance $d$ are $\sim10\%$. The selection functions used are that of the DESI Milky Way Survey and of Gaia (see Sec.~\ref{sec:obs_mocks}). The uncorrected $\sigma_t^2$ profiles deviate strongly from the true profiles beginning around $r=40$~kpc but are very well recovered by deconvolution. The deconvolution routine has varying success when recovering the $\rho$ profiles. It is done excellently for \textit{m12f}, but quite poorly for \textit{m12i} and \textit{m12m} within $r=10$~kpc.}
        \label{fig:deconv_sig&dens}
    \end{center}
\end{figure*}

All of our previous tests have assumed that the data available to us are error-free, full 6D phase space coordinates of halo tracers throughout the volume of interest. This simplification enabled us to assess the systematic error introduced by the code itself, but did it not allows us to assess how realistic observational selection functions and errors affect the Jeans mass estimates. In reality, practically all observations come with errors and spatial selection functions. In this section, we describe tests on the observationally motivated mock datasets described in Section~\ref{sec:obs_mocks} and the more advanced analysis we apply to them described in Section~\ref{sec:obs-deconv}. These mocks better represent the data we expect to obtain from Gaia and DESI by imposing their observational selection functions and uncertainties in phase space coordinates of tracers. 
We now model the density and velocity distributions in the halo by deconvolving the observational effects of incomplete and imprecise data from the true underlying data. Figure~\ref{fig:deconv_sig&dens} shows example output of our deconvolution routine on the three Latte galaxies in question with mock Gaia (E)DR3 and DESI errors imposed at LSR0. The MCMC fits (solid colored lines) recover the true density and velocity dispersion profiles (solid black lines) despite the input data (dashed colored lines) deviating from them significantly. In the right column the proper motion errors on the data clearly blow-up at large radii for the observed tangential velocity dispersion profile (dashed blue line), but the effect is entirely corrected for by deconvolution in all three galaxies (solid blue line). Properly accounting for the selection function also largely mitigates the effect of the incomplete spatial coverage of the survey on the tracer density profile, as seen most clearly as a sharp drop-off in density (dashed red curve) at $r<10$ kpc in the left column. The accuracy of the corrected $\rho$ profile is not uniform, however. While \textit{m12f} has its density well recovered by the deconvolution routine, \textit{m12i} and \textit{m12m} still have large errors within $r=10$~kpc. We attribute this to their elevated amounts of substructure and streaming motion as seen in Figure~\ref{fig:heatmaps}. We also note that the true density profile and error-imposed density (red dashed curve)  of \textit{m12m} line up nearly exactly beyond $r=80$~kpc. This is undoubtedly the result of an enormous stream, visible in Fig.\ref{fig:heatmaps}, dominating the halo density at that radius. We discuss \textit{m12m} in light of this stream and its relevance to our future MW application later in this section.

\begin{figure*}
    \begin{center}
        \includegraphics[trim=0 0 0 0,clip,height=11cm,keepaspectratio]{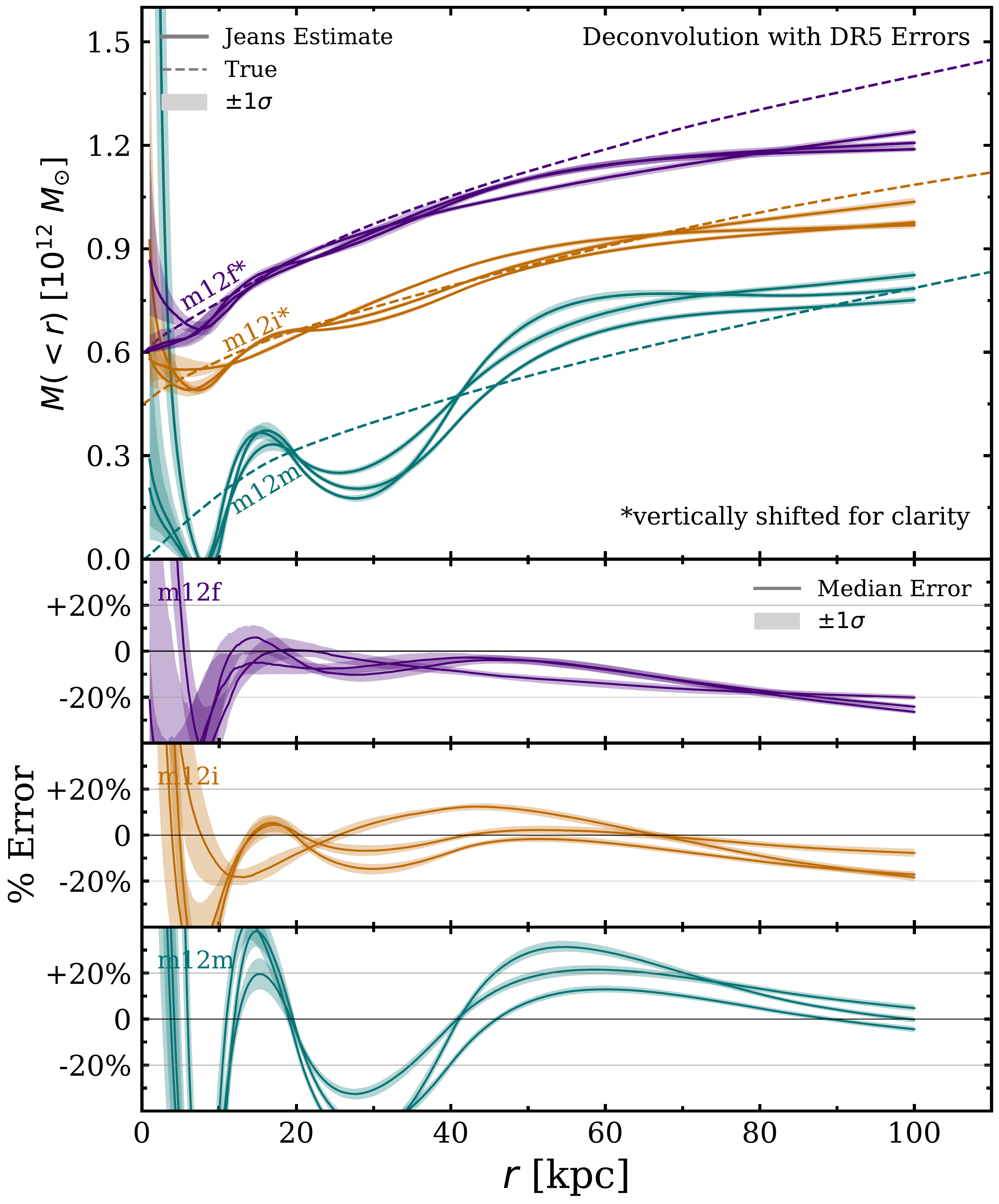}
        \includegraphics[trim=133 0 0 0,clip,height=11cm,keepaspectratio]{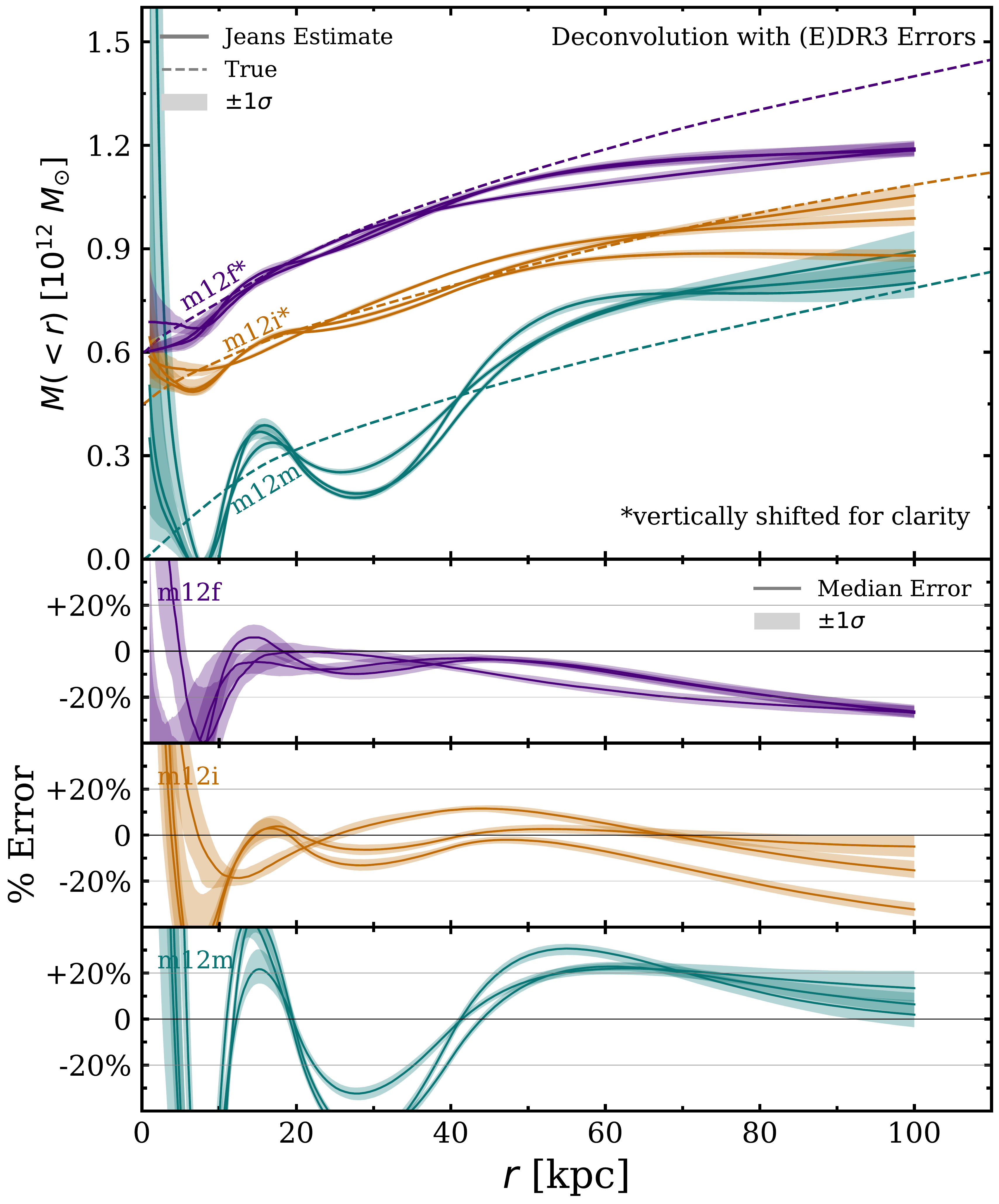}
        \caption{Results of Jeans estimation on Latte galaxies \textit{m12f} (purple), \textit{m12i} (orange), and \textit{m12m} (teal) after deconvolving observational effects from underlying data. Mocks with Gaia DR5 proper motion errors are shown in the left column and (E)DR3 proper motion errors in the right column. The bands above and below each Jeans mass estimate represent the 16--84 percentile range of the models in the MCMC series. The three lower panels show the errors on the Jeans mass estimates for the three galaxies. The accuracy at $r=100$ kpc is usually within 20\% for either Gaia release, but the uncertainty bands and the spread between LSRs are notably larger for the (E)DR3 mocks than for the DR5 mocks}
        \label{fig:deconv_fig}
    \end{center}
\end{figure*}

We can now plug the deconvolved density and velocity profiles into the spherical Jeans equation (Eq.~\ref{sph_jeans_eq}) to calculate the enclosed mass profile $M(<r)$. Figure~\ref{fig:deconv_fig} shows results of the deconvolution run on Latte galaxies {\it m12f, m12i}, and {\it m12m} with mock Gaia DR5 and (E)DR3 errors. As described in Section~\ref{sec:obs_mocks}, we consider three LSRs for each galaxy whose respective mass profiles are distinguished as separate solid colored lines. The colored uncertainty bands above and below the estimated mass profiles are 16--84 percentile intervals of models in the MCMC series. 

From Figure~\ref{fig:deconv_fig}, there are a couple key takeaways. In the DR5 panel (left), the small radii region ($r<20$~kpc) has large errors and dramatic swings between over- and underestimating the enclosed mass across all three galaxies. This is not particularly surprising as much of this region had all the star particles removed by the observational selection functions, so we expect low confidence in this region. Beyond about $r=20$~kpc, the mass profiles for \textit{m12f} and \textit{m12i} are stable and yield impressive accuracy. It is only at about $r=85$~kpc when some of the \textit{m12f} mass profiles reach more than 20\% error while all the \textit{m12i} mass profiles stay within 20\% out to $r=100$~kpc. The results are quite different, however, for \textit{m12m}. After the dramatic swings within 20~kpc, the \textit{m12m} mass profiles continue exhibiting oscillatory motion around their true mass profile. Such poor performance is expected considering the tremendous streaming motion in \textit{m12m} (which we ignored in our deconvolution algorithm) in comparison to {\it m12f} and {\it m12i} (Fig.~\ref{fig:heatmaps}). We note that even with perfect data over the entire volume \textit{m12m} showed the largest deviation from the true mass distribution (see Fig.~\ref{fig:latte_fig}), but since the formulation of the Jeans mass estimate with perfect data shown in Fig.~\ref{fig:latte_fig} did not assume $\overline{v}_{r,\phi,\theta}=0$, the results were better than the deconvolved estimates which assume no streaming motion. In future applications, we may need to add more flexibility in the models by allowing a nonzero streaming motion around a particular axis (not necessarily perpendicular to the disc plane). It must be noted, however, that \textit{m12m} should be considered a worst-case-scenario. When we progress to real data we will not encounter many of the reasons that \textit{m12m} produced a poor fit. There is not nearly as much streaming motion in the Milky Way as there is in \textit{m12m}, and the well defined streams that are present, such as Sagittarius, are often easily removed from observational datasets. For these reasons, we expect results using MW data to be more akin to that from \textit{m12f} and \textit{m12i} rather than \textit{m12m}.

We emphasize here that while non-monotonic cumulative mass profiles are completely unphysical they arise from the fact that the enclosed mass estimate at a given radius depends only on the tracer density profile and velocity distribution at that radius. We did not impose any relation between density and velocity dispersion profiles in the model from the outset to preclude non-monotonic mass profiles. It is straightforward to add this constraint in the fitting procedure (disqualifying non-monotonic models from the MCMC chain), but it is unlikely to mitigate the underlying physical reasons for the poor fit.

In the (E)DR3 panel (right), the results are remarkably similar to the DR5 panel (left) despite the much greater effect of larger proper motion errors on pre-deconvolution data (Fig.~\ref{fig:gaia_drs}). The \textit{m12f} results are nearly identical across (E)DR3 and DR5 runs. The \textit{m12i} results are also nearly identical until about $r=50$~kpc, beyond where the spread between different LSRs and the uncertainty bands are much greater for the (E)DR3 than the DR5 runs. The (E)DR3 \textit{m12m} results have the same issues as those of DR5 but, similarly to \textit{m12i}, with increased spread between LSRs and increased uncertainty band size. The similarity in accuracy is a testament to the efficacy of the deconvolution routine when dealing with large or small proper motion errors.

\section{Summary \& Conclusion}
\label{sec:conclude}

We have presented a new non-parametric spherical Jeans modeling code that uses B-splines to fit the enclosed mass distribution within the Milky Way halo. We present tests of this code with a variety of mock datasets to assess the effects of breaking various assumptions made in deriving the spherical Jeans equations. Our main results are listed below:
\begin{itemize}
\item With our simplest ``Halo-alone'' (HA) models we show that going from a spherical mass distribution to a flattened mass distribution with axis ratio $q=0.6$ only slightly increases the error on the estimate of enclosed mass but the error is within 4\% beyond $r=5$ kpc (see Fig.~\ref{fig:HA_fig}).

\item Our ``Halo-Disc-bulge'' (HDB) models are constructed to resemble the Milky Way halo, disc, and bulge but consider only a spherical halo. We also include the additional test of considering both isotropic and anisotropic velocity distributions for halo particles, and the test of having ``contamination'' from disc stars in the halo. Our Jeans modeling code performs very well with these HDB models with maximum errors on the enclosed mass almost always within 10\% and less than 5\% at most radii (see Fig.~\ref{fig:HDB_fig}). 

\item The mock datasets with the next level of complexity are constructed from the Latte FIRE-2 zoom-in cosmological hydrodynamical simulations of three galaxies, \textit{m12f, m12i}, and \textit{m12m}. These galaxies were considered Milky Way analogs and contain realistic amounts of substructure and display varying amounts of disequilibrium (see Fig.~\ref{fig:heatmaps}). With error-free data over the whole sky, the code still gives better than 15\% error out to 100~kpc in all three Latte galaxies (see Fig.~\ref{fig:latte_fig}). 

\item The most realistic datasets build upon the previous by considering the effects of observational selection functions and errors from Gaia and the DESI Milky Way Survey \citep{DESI-MWS_RNAAS_2020}. The Gaia component includes a $G$-band magnitude upper-limit, proper motion errors from the end of (E)DR3 and DR5, and errors on spectro-photometric distances. The DESI component includes its observational footprint (declination and galactic latitude bounds) and errors on line-of-sight velocity. These errors and selection functions are imposed on the same three Latte (FIRE-2) galaxies and consider three possible solar positions. Paired with the addition of observational effects is the inclusion of a more advanced method for modeling the velocity and density profiles. This method is designed to deconvolve the effects resulting from the observational errors and selection functions causing the data to be imprecise and incomplete. The two primary observational effects are on the density at small Galactocentric radii, due to the removal of volume by the DESI footprint, and on the tangential velocity dispersion at large radii, due to Gaia proper motion errors scaling up with heliocentric distance. Figure~\ref{fig:deconv_sig&dens} shows the $\sigma_t^2$ and $\rho$ profiles of the underlying incomplete and imprecise dataset as well as their corrected counterparts after applying deconvolution. Jeans estimates performed with this setup yield very encouraging results (see Fig.~\ref{fig:deconv_fig}). When run with DR5 proper motion errors, the error on $M(<100 \textrm{ kpc})$ is around 20\% and often better. The (E)DR3 runs give very similar results for \textit{m12f} but have greater spread between LSRs in \textit{m12i}, and \textit{m12m}. The resulting mass profile for \textit{m12m} has dramatic oscillations above and below true mass profile and violates monotonicity for cumulative mass profiles. We expect this breakdown of our Jeans routine is due to the very large streaming motions present in \textit{m12m} which is not accounted for in our deconvolution routine (see Fig.~\ref{fig:heatmaps}).


\end{itemize}

The spherical Jeans equation method is a fairly simple, but popular method for recovering the mass distribution of (nearly) spherical potentials. A fundamental feature of this method is that it depends on the radial derivatives of the tracer mass distribution as well as the radial velocity dispersion profile. Since derivatives of observational data profiles are extremely noisy \citep{kafle_etal_18}, most authors model the tracer density and velocity dispersion profiles with analytic functions (often power-laws) to enable convenient computation of the derivatives. In our implementation of this method, we have used B-splines to estimate the density profile and velocity dispersion profiles of the tracer particles, enabling us to avoid the need to assume a functional form but still allowing for the rapid computation of the enclosed mass profile\footnote{The routine's run time without deconvolution is a few seconds.}. 

Other implementations of the spherical Jeans equation formalism applied to full 6D phase space data \citep[e.g.][]{wang_etal_18} have also been tested on several hundred isolated and binary haloes from the dark-matter only Millenium II simulations \citep{Boylan-Kolchin_etal_2009} and 12 intermediate resolution cosmological hydrodynamical zoom-in simulations of Local Group analogues  \citep[the APOSTLE simulations,][]{Fattahi_etal_16, Sawala_etal_16}. \citet{wang_etal_18} find that their Jeans estimates of $M_{200}$ and the halo concentration parameter $c_{200}$ yielded errors of around 25\% when dark matter particles are used as tracers but the errors can go up to 200-300\% when using star particles as tracers (even with error-free phase-space data). The APOSTLE datasets used by these authors are most directly comparable to our error-free Latte datasets (Sec.~\ref{sec:latte_mocks}), on which we achieve errors on $M(<100\textrm{ kpc})$ of $\sim5-15\%$ (Fig.~\ref{fig:latte_fig}). We believe this discrepancy is likely due to three factors. First, although not explicitly stated in their paper we infer that \cite{wang_etal_18} used binned data for the density $\rho(r)$, velocity dispersion $v_r(r)$ and anisotropy $\beta(r)$ profiles which were likely quite noisy. Our early experiments with binned data showed that numerical derivatives of such data are extremely noisy and do not have the desirable qualities B-splines provide our implementation: smooth non-parametric fits and analytical derivatives. Second, \citet{wang_etal_18} cut out the inner 20~kpc to exclude the \textit{in situ} stellar component but include tracers out to the virial radius. We instead selected old metal poor stars with $\lbrack\mathrm{M}/\mathrm{H}\rbrack<-1.5$ and $t_\textrm{age}>8$~Gyr in an attempt to capture a more virialized halo population and placed our maximum knot at $r=80$~kpc, within the $\sim100$~kpc limit  where we expect a significant sample of Gaia+DESI data. Our experiments selecting more metal rich stars do show a small increase in the estimated errors on $M(<100\textrm{ kpc})$ but only up to $20\%$ (not 300\%). Increasing the radius of our maximum knot to 100~kpc and evaluating our B-splines out to 200~kpc with extrapolation maintains our great accuracy at 15\% error on $M(<200 \textrm{ kpc})$. Finally, while \citep{wang_etal_18} fit their derived potentials to NFW potentials and report the errors on the best-fit $M_{200}, c_{200}$ we simply report the errors on the actual cumulative mass profiles of our simulated haloes, which we do not compare with NFW profiles. While other factors like the specific implementation of baryonic microphysics in the different simulations might cause some differences, the three galaxies from the Latte suite of FIRE-2 simulations we have chosen in this study have global properties quite similar to the Milky Way. Our tests in this article suggest that, for real Gaia+DESI data, the prospects for obtaining errors of no more than 25\% are quite favorable and would be a significant improvement over the current factor of 2 uncertainty in halo mass.

Other methods such as distribution function fitting \citep{Watkins_etal_2010, eadie_etal_17, eadie_etal_18, eadie_etal_19, hattori_etal_21, deason_etal_21, Shen_J_etal_2021} use Bayesian MCMC analysis to properly account for the errors on the mass estimates arising from the error on the data. These methods, while much more sophisticated, are computationally and technically challenging, and for simplicity, most authors assumed very simple power-law profiles for the tracer density and the potential. Our goal in developing this non-parametric B-spline Jeans modeling code and testing it with mock data is to provide the community with a fast and easy-to-use code with well quantified systematic errors. We dub the code \textsc{nimble} (Non-parametrIc jeans Modeling with B-spLinEs) and have made it publicly available on GitHub (\url{https://github.com/nabeelre/nimble}). Included in the repository are the routines for performing the analysis described in Sec.~\ref{sec:precise-complete} and Sec.~\ref{sec:obs-deconv} as well as code to generate all the mock datasets described in Sec.~\ref{sec:mockdata}. Generally, it can be applied to a sample of observed 6D coordinates for tracer stars to construct B-spline estimates of the radial density profile, velocity dispersion profiles and the velocity anisotropy profile, which are then used to evaluate Eq.~\ref{sph_jeans_eq}.

In future, we aim to apply this code to Gaia DR4 and Gaia DR5 data with line-of-sight velocities from DESI, as well as future data expected from WEAVE \citep{weave} and 4MOST \citep{4MOST}. 

\section{Data Availability}
The data in this article come from two primary sources. 

The data described in Sec.~\ref{sec:agama_mocks} are generated using the \textsc{Agama} package \citep{vasiliev_19_agama} using scripts provided in the \textsc{NIMBLE} repository, which can be accessed at \url{https://github.com/nabeelre/nimble}. 

The data used to make the mocks described in Sec.~\ref{sec:latte_mocks} and \ref{sec:obs_mocks} are of the Latte suite of FIRE-2 cosmological zoom-in simulations \citep{wetzel_etal_16} made publicly available and provided by the FIRE collaboration under Creative Commons BY 4.0. 

\section{Acknowledgements}
We thank the anonymous referee for their valued comments and members of the stellar haloes group at the University of Michigan for stimulating discussion. We especially thank Eric Bell and Kohei Hattori for assistance and advice. M.V. and N.R. were supported in part by NASA grants NNX15AK79G and 80NSSC20K0509 and a Catalyst Grant from the University of Michigan's Michigan Center for Computational Discovery and Engineering (MICDE).

In addition to \textsc{Agama} \citep{vasiliev_19_agama} we used the following software packages: Astropy \citep{astropy:2013, astropy:2018}, emcee \citep{emcee}, Matplotlib \citep{hunter07}, NumPy \citep{numpy}, SciPy \citep{scipy}, and PyGaia (\url{https://github.com/agabrown/PyGaia})

\bibliographystyle{mnras}
\bibliography{halorefs}

\appendix
\section{Fitting Density and velocity profiles with B-splines}
\label{sec:bsplines}
In this section we provide a brief overview of the B-spline formalism in the context of fitting and density estimation. For a more complete mathematical background see Appendix A2 in \citet{AgamaReference}.

B-splines are a set of piecewise-polynomial basis functions $B_j(x)$ defined by knots of a grid $x_k, k=1..K$. B-splines of polynomial degree $N$ are nonzero on at most $N+1$ consecutive grid segments, their $N-1$'th derivatives are continuous at grid knots, and a grid with $K$ knots ($K-1$ segments) generates $K+N-1$ basis functions. A B-spline representation of a curve is a linear combination of basis functions with some coefficients: $f(x) = \sum_j A_j\, B_j(x)$. The most familiar case is a histogram, which is none other than a 0th degree B-splines basis; however, since even the curve itself is discontinuous, it is not suitable for evaluating derivatives. In this study we use 3rd degree B-splines, which are mathematically equivalent to a clamped cubic spline curve (clamped meaning that its 2nd derivatives at endpoints are not necessarily zero). 

In the present context, we use B-splines to estimate the density from discrete samples. Namely, given a set of points $x_n$ with weights $w_n$, $n=1\dots N_\mathrm{data}$, we seek to construct a density function $p(x)$ such that $\int_a^b p(x)\,dx \approx \sum_n w_n\,\sqcap(x_n;\, a,b)$, where the indicator function $\sqcap(x;\, a,b)=1$ when $a\le x \le b$ and zero otherwise. The density is normalized so that the integral over the entire interval containing all points equals $\sum_n w_n$. It turns out that B-splines are very useful for representing the logarithm of $p$, ensuring the positivity constraint on density. After choosing the interval and grid knots, the coefficients $A_j$ of the B-spline representation $\ln p(x) = \sum_j A_j B_j(x)$ can be found by maximizing the penalized log-likelihood of the input sample: $\{A_j\} = \mathrm{arg\,max} \sum_n w_n\,\ln p(x_n) - \lambda\int \big[d^2\,\ln p(x) / dx^2\big]^2\, dx$, where the second term biases the solution towards smooth curves, and the roughness penalty $\lambda$ is automatically determined by cross-validation. With the optimal choice of $\lambda$, the number of grid knots has little effect on the resulting curve (as long as it is sufficient to resolve key features in the distribution). Note that since the density is normalized to unity, the number of free parameters in the fit is one fewer than the number of spline knots.

For the Jeans equation, we need the density and the kinetic energy, which are estimated as follows. The distribution of points in spherical radius (the number of points in the radial interval $r..r+dr$) is $dN(<r)/dr = 4\pi\, r^2\,\rho(r)$, and it is convenient to use $\chi\equiv \ln r$ instead of radius. We construct a B-spline estimate for the logarithm of the density of points in $\chi$, using the masses $m_n$ of particles in the input snapshot as the weights associated with points: $f_1(\chi) \equiv \ln \big[ dN / d \ln r \big] = \ln \big[ 4\pi\,r^3\,\rho(r) \big]$; hence the mass density is given by $\rho(r) = (4\pi)^{-1}\, \exp\big[ f_1(\chi) - 3\chi \big]$.
Likewise, when using the masses multiplied by squared radial and tangential velocity components $v_r^2$, $v_\theta^2+v_\phi^2$ of each particle as the weights, we obtain the estimates of the corresponding terms in the kinetic energy tensor: $f_{v_r^2}(\chi) \equiv \ln \big[ 4\pi\,r^3\,\rho(r)\,\overline{v_r^2} \big]$, etc. The logarithmic derivative of the radial kinetic energy $d\ln \big[\rho\,\overline{v_r^2}\big] / d\ln r$ appears in the Jeans equation directly, and the other terms contains the ratios of $\exp f_1$, $\exp f_{v_r^2}$, etc.


\bsp	
\label{lastpage}
\end{document}